\documentclass[showpacs, aps, prb, unsortedaddress,showkeys,longbibliography]{revtex4-1}

\usepackage{amssymb}
\usepackage{latexsym}
\usepackage{amsmath}
\usepackage{graphicx}
\usepackage{braket}
\usepackage{multirow}

\usepackage[dvipsnames]{xcolor}
\usepackage{float}
\begin{document}

\title{Scattering of mechanical waves from the perspective of open systems}

\date{\today}
\author{Hossein Khodavirdi}
\author{Amir Ashkan Mokhtari}
\author{Ankit Srivastava}
\thanks{Corresponding Author}
\email{asriva13@iit.edu}

\affiliation{Department of Mechanical, Materials, and Aerospace Engineering
Illinois Institute of Technology, Chicago, IL, 60616
USA}

\begin{abstract}
In this paper, we consider the problem of mechanical wave scattering from a spatially finite system into an infinite surrounding environment. The goal is to illuminate why the scattering spectrum undergoes peaks and dips (resonances) at specific locations and how these locations connect to the vibrational properties of the scatterer. The resonance locations are connected to the eigenvalues of a finite dimensional effective operator, $H_{eff}$, corresponding to the scatterer. The developments are presented from the perspective of open systems, which seeks to convert the infinite dimensional scattering problem (scatterer+environment) into a finite dimensional effective problem involving only the finite scatterer. This is achieved through a projection operator formalism which allows us to formally calculate $H_{eff}$. An interesting corollary of our analysis is the deep connection between resonance locations in the scattering spectrum and the eigenfrequencies of the scatterer under Neumann boundary condition. We bring out this point further by considering 3D scattering from an elastic shell, connecting our results to classical results in acousto-elastic scattering theory. 


\end{abstract}

\maketitle

\section{Introduction}

In this paper, we are concerned with understanding the scattering of mechanical waves from scatterers of finite extent. This problem has a long history of research in the general area of mechanics beginning from classic works~\cite{rayleigh1917reflection,King,willis1980polarization,eshelby1957determination,westervelt1957acoustic,faran1951sound,krein1956theory,krein1962theory}. More recently, scattering problems have also been of interest in cloaking problems~\cite{milton2006cloaking,wolf1993invisible,chen2007acoustic,cummer2008scattering,norris2008acoustic}, in problems involving metasurfaces~\cite{Krasnok_2018,assouar2018acoustic,chen2016review}, in those concerning nanoscale thermal transport \cite{Nanoscalethermalreview,AGFNanoscalePhonon,PhysRevB.91.174302}, and other related phononic and metamaterial problems\cite{MHusseinReview,mokhtari2020scattering,mokhtari2019properties,baz2010active}. In general, the concern with scattering problems is to calculate the spectrum of scattering (scattering amplitude as a function of frequency) given a scatterer. There are numerous techniques of doing so~\cite{skudrzyk2012foundations,roach2009wave}. Here we ask a related question - what is the relation between the prominent features of the scattering spectrum (locations of peaks and dips) and the vibrational eigenfrequencies of the scatterer? The main theoretical machinery is that of open quantum systems which is used to study the scattering of quantum waves from potentials localized in space. Conservation of probability is a key feature in certain systems in quantum mechanics and it is related to the real nature of the eigenvalues of the Hermitian Hamiltonians of such systems~\cite{NonHermitianPhys,fanobook}. For example, for the classical particle in a box case, one can calculate the probability of finding the particle at a given location inside the box as the square of the wave function. Decreasing the probability of finding the particle at a position is followed by increasing the probability of finding it in another position -- and the total probability of finding the particle in the box sums to 1~\cite{cohen2006quantum,kamenetskii2018fano} -- this is the conservation of probability. However, this probability is not conserved when a quantum system (such as the box) is placed in an environment with which it can interact. In such a configuration, the particle can escape the box to infinity and the probability of finding it inside the box does not sum to 1. Such systems, where an otherwise conservative finite system is coupled with an environment, thus giving rise to the flows of energy, particle, or information between the finite system and the environment are called open systems~\cite{Livsic}. An open system, in a general sense, consists of a finite subsystem with discrete eigenvalues (levels of energy) which is coupled to an environment possessing a spectrum which is continuous~\cite{boundscattering}. Since these systems are inherently non-conservative, their effective Hamiltonian operators are non-Hermitian, giving rise to complex eigenvalues. An equivalence between the single-particle Schrodinger equation in quantum mechanics and classical wave equation makes it possible to study the scattering of classical waves from the point of view of open systems and the associated non-Hermitian effective Hamiltonian problem. 

The deep connection between scattering problems and open system problems also means that the open system effective Hamiltonian can be useful for certain scattering calculations. For example, while it is sometimes difficult to find the poles of the scattering matrix using the standard Hermitian formalism, it can be done directly by calculating the eigenvalues of the associated non-Hermitian Hamiltonian of the corresponding open system~\cite{moiseyev2011non}. Non-Hermiticity can give rise to unconventional phenomena in wave transport such as unidirectional light reflection in PT-symmetric systems~\cite{unidirectionalref} and enhanced sensitivity to external perturbations~\cite{LinearPerturbation,Sensitivityanalysis,multipleeigenvalue}. Additionally, parametric non-Hermitian systems have also been of interest to the research community~\cite{heiss2012physics,LinearPerturbation,heiss2001chirality,heiss2004exceptional}. This has revealed exotic phenomena like level repulsion and exceptional points \cite{LU2018100,heiss2012physics,Avoidedlevelcrossings}, non-orthogonality of eigen-modes \cite{Nonorthogonalityconstraints}, self-orthogonality in certain systems~\cite{heiss2001chirality}, high sensitivity to parametric variation \cite{LinearPerturbation,Sensitivityanalysis,multipleeigenvalue,hodaei2017enhanced,chen2017exceptional}, and chirality of the eigenstates~\cite{heiss2001chirality} near the exceptional points.

Non-Hermiticity has also been studied in the field of metamaterials. Metamaterials have been proposed which exhibit effective properties with non-zero imaginary parts, implying either material gain or loss exists in the system~\cite{frazier2016generalized, mokhtari2019properties}. Such metamaterials allow for more design degrees of freedom, made possible by exploiting the interplay between loss, gain, and the coupling between individual resonances~\cite{zangeneh2019topological}. Acoustic gain in such metamaterials can be achieved through various methods, such as by using piezoelectric materials \cite{Willatzen2014}, electroacoustic circuits \cite{Fleury2015}, exploiting the coupling between sound and hydrodynamic instabilities \cite{Aurgan2017}, or thermo-acoustic effects\cite{thermoacoustic}. Non-hermiticity also appears in metamaterials involving  parity ($P$) and time ($T$) symmetries. An interesting point about such systems is that in spite of having a non-Hermitian Hamiltonian, they have real eigenvalue spectra below some threshold related to an exceptional point. Within this regime, the losses are compensated by the gain, providing loss-immune acoustic metamaterials. Above the threshold, the system enters the $PT$-symmetry broken phase with complex conjugate eigenvalues \cite{Achilleos2017,Fleury2014, ElGanainy2018, Monticone2016, Konotop2016}. These extraordinary spectral features lead to exceptional scattering behavior such as anisotropic transmission resonances, also called unidirectional cloaking \cite{Fleury2015}, and coherent perfect absorbers and lasers \cite{Song2014, Wei2014}. 

An important and useful tool in analyzing systems which interact with their environment is the \textit{effective Hamiltonian}. Effective Hamiltonian is the Hamiltonian calculated for only a part of the problem and it considers the effect of the rest of the problem implicitly. It acts in a reduced space and only describes a part of the eigenvalue spectrum of the true (more complete) Hamiltonian. An important technique of formally calculating the effective Hamiltonian of an open quantum system is using projection operators like the Feshbach projection operator formalism~\cite{FESHBACH1958357,Feshbach1962}. A useful application of finding the effective Hamiltonian of an open system is solving scattering problems once an incoming wave from the environment is incident at the subsystem of interest. We show how Green's function can be derived from the effective Hamiltonian and then how scattering coefficients can be calculated using Green's function. In earlier works~\cite{Deymier2017}, some scattering problems in discrete mass-spring systems are solved using the interface response theory. This formalism allows the calculation of the Green 's function of a perturbed system in terms of the Green's functions of unperturbed systems. Although both methods are based on the calculation of Green's functions, the method of interface response theory follows a complicated and relatively long path. On the other hand, using the open system point of view and calculating the effective Hamiltonian present a simpler and physically more understandable approach in solving wave scattering problems. \par

The methods followed here were developed in the area of nanoscale heat transfer~\cite{sadasivam2014atomistic,mingo2003phonon,khalatnikov2018heat} where various traditional methods of calculating scattering in low/high temperature regimes and for ideal/non-ideal interface surface roughness have emerged \cite{little1959transport,swartz1989thermal} in the past. The method used in this paper is connected to the Atomistic Green’s function method (AGF), which stemmed from the theory of coherent transport of electrons \cite{datta2005quantum}. AGF is used mostly for harmonic phonon processes and for simplified interfacial models, such as junctions \cite{nano} or nanowires \cite{mingo2003phonon}, and it allows for the calculation of the effective Hamiltonian and scattering parameters~\cite{sadasivam2014atomistic}. The related efforts in nanoscale heat transfer have gone further and an extended version of the AGF method, using the concept of Bloch matrix \cite{PhysRevB.72.035450,PhysRevB.44.8017}, determines not only the total transmission function but also the transmission due to each single phonon modes \cite{PhysRevB.91.174302}.

In this paper we use a 1D discrete mass-spring model, similar to tight-binding systems in quantum mechanics, to show how the concepts of effective Hamiltonian and projection operator formalism can be applied to a mechanical wave propagation problem. First we present an example of a scattering problem and simply point out a few interesting observations about the scattering spectrum in it. The main question concerns the understanding of the connection between the vibrational features of the scatterer and the features in the scattering spectrum of a wave scattered by the scatterer. Section \ref{Feshbachintro} presents a short introduction to a projection operator formalism and Section \ref{section5} explicitly shows that, for the problem under consideration, the effective Hamiltonian derived from the projection formalism is exactly the same as that derived from a direct method. This section also elucidates the connection between the eigenvalues of the effective Hamiltonian, the Green's function, and the scattering spectrum. Finally, a scattering problem for a 4-DOF scatterer is the basis for several numerical examples in section \ref{NR}, where we discuss the effects of the couplings inside/outside of the scatterer on the features of the scattering spectrum. We also show that some of the lessons learned from the projection operator formalism can be used to understand resonance scattering in higher dimensional problems such as scattering from a spherical shell. Such problems have a long history in mechanics where the method of choice for understanding the resonant features in the scattering spectrum is the resonance formalism~\cite{faran1951sound}. Flax et. al. \cite{flax1978theory} considered the problem in detail and showed that the scattering resonances are caused by the superposition of resonances in the individual normal mode amplitudes of the scatterer (see also \cite{neubauer1969theory,uberall1977relation,hackman1989existence,hackman1993acoustic,klaiman2010absolute}).

\section{A scattering problem}\label{section2}

\begin{figure}[htp]
\centering
\includegraphics[scale=.33]{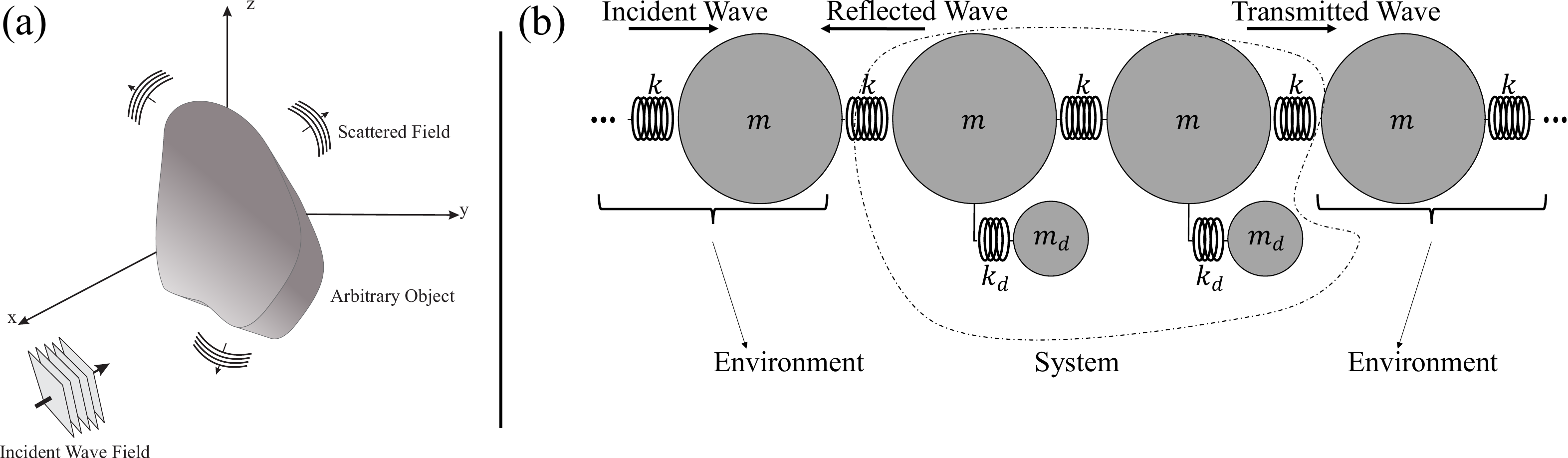}
\caption{(a) A general scattering problem. (b) An illustration of a simple discrete open system. Two semi-infinite chains of masses and springs connected to a central system with resonators.}
\label{tworesonators}
\end{figure}

Consider the scattering problem in Fig. (\ref{tworesonators}). Fig. (\ref{tworesonators}a) shows a conceptual schematic of an open system. It consists of a heterogeneity (scatterer) of a finite extent in an otherwise infinite homogeneous problem. The finite heterogeneity is called the system whereas the homogeneous medium is called the environment. Fig. (\ref{tworesonators}b) shows a specific manifestation of this general problem in the context of a 1-D discrete chain of masses and springs. The problem consists of a finite central region of masses and springs (the system) connected to two semi-infinite homogeneous mass-spring chains (the environment). In either of these problems, waves traveling in the environment are scattered by the system. The scattering spectrum in the environment is critically connected to the properties of the system, a connection which is encapsulated in the effective Hamiltonian.

For Fig. (\ref{tworesonators}b), we consider a specific case where $k=1$ N/m, $m=m_d=1$ kg, and the spring constants in the left and right resonators are $k_{d_0}=1.01$ N/m and $k_{d_{1}}=0.8$ N/m respectively. We now assume that an incident wave travels in the environment in, and gets scattered (reflected and transmitted) by the system. The incident and scattered waves are all characterized by a real frequency $\omega$. Waves which are traveling to the right have a wavenumber $q$ and those traveling to the left have the wavenumber $-q$. $\omega,q$ are related to each other through the dispersion relation of the environment. In this case, since the environment is simply a homogeneous chain of mass and spring, the dispersion relation has a well known analytic form~\cite{kittel1976introduction}. For the current problem with the above parameters, the $q$ solution for $0\leq\omega\leq 2$ is real. We call these real $q$ wave solutions propagating modes, but they are also referred to as continuum states in quantum mechanical parlance. Purely imaginary $q$ solutions are called evanescent waves or bound states. Finally, complex $q$ solutions comprise the resonant ($\Re q>0,\Im q<0$) and anti resonant ($\Re q<0,\Im q<0$) states. Thus, we have a full picture of the wavenumber in the complex $q$ domain. Similarly, we will consider the complex frequency domain as well. The incident and scattered waves in this paper will always be assumed to have a real frequency but the eigenvalues of the effective Hamiltonian will necessarily be in the complex frequency domain.

\begin{figure}[htp]
\centering
\includegraphics[scale=0.182]{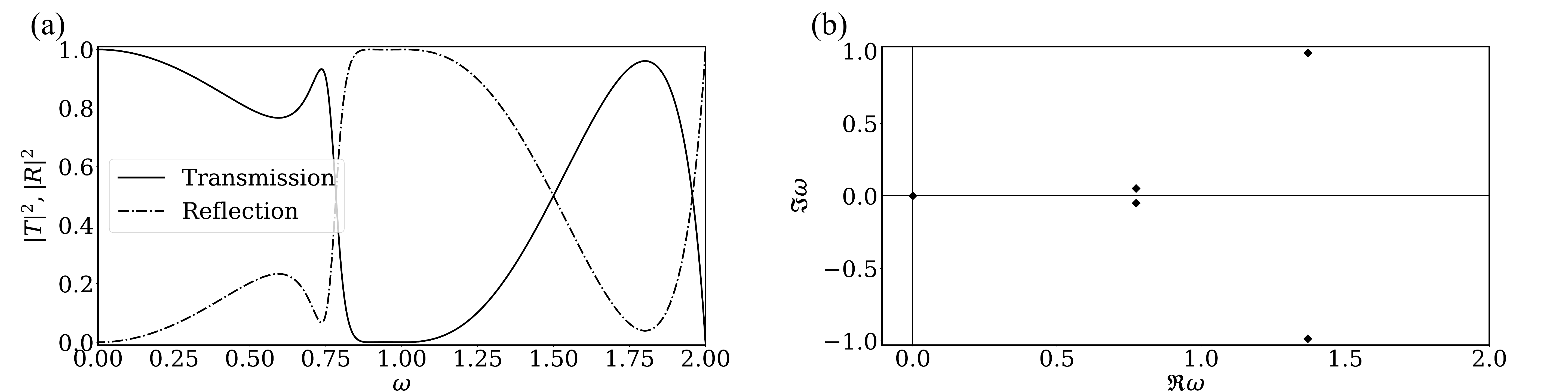}
\caption{  (a) The transmission and reflection vs. angular frequency are shown. (b) The Imaginary vs. real parts of discrete poles are shown.}\label{Transmissionfg}
\end{figure}

We now assume that incident wave is of unit amplitude. In this simple problem, the reflected and transmitted waves in the environment have the same frequency and magnitude of the wavenumber as the incident wave, and their amplitudes are given by complex constants $R,T$ respectively. Fig. (\ref{Transmissionfg}) shows $|T|^2(\omega),|R|^2(\omega)$ in the frequency range where a propagating wave solution exists in the environment ($q$ is real). The equations needed for these and related calculations will be provided in the subsequent sections. We note here that the scattering spectrum for this simple problem exhibits certain peaks and valleys. The salient peaks of the transmission spectrum exist at around $\bar{\omega}_i= 0.74$ and $1.79$ and there are corresponding valleys for the reflection spectrum at these locations. These locations are highlighted in the figure through dashed lines. Here, we are interested in two main questions: \emph{why do the peaks and dips of the scattering spectrum appear at the precise locations where they are occurring and how are these locations related to the vibrational properties of the system?}

\subsection{Relation between $\bar{\omega}_i,\omega_i,{\omega}_i^N,{\omega}_i^D$}

\begin{figure}[htp]
\centering
\includegraphics[scale=.33]{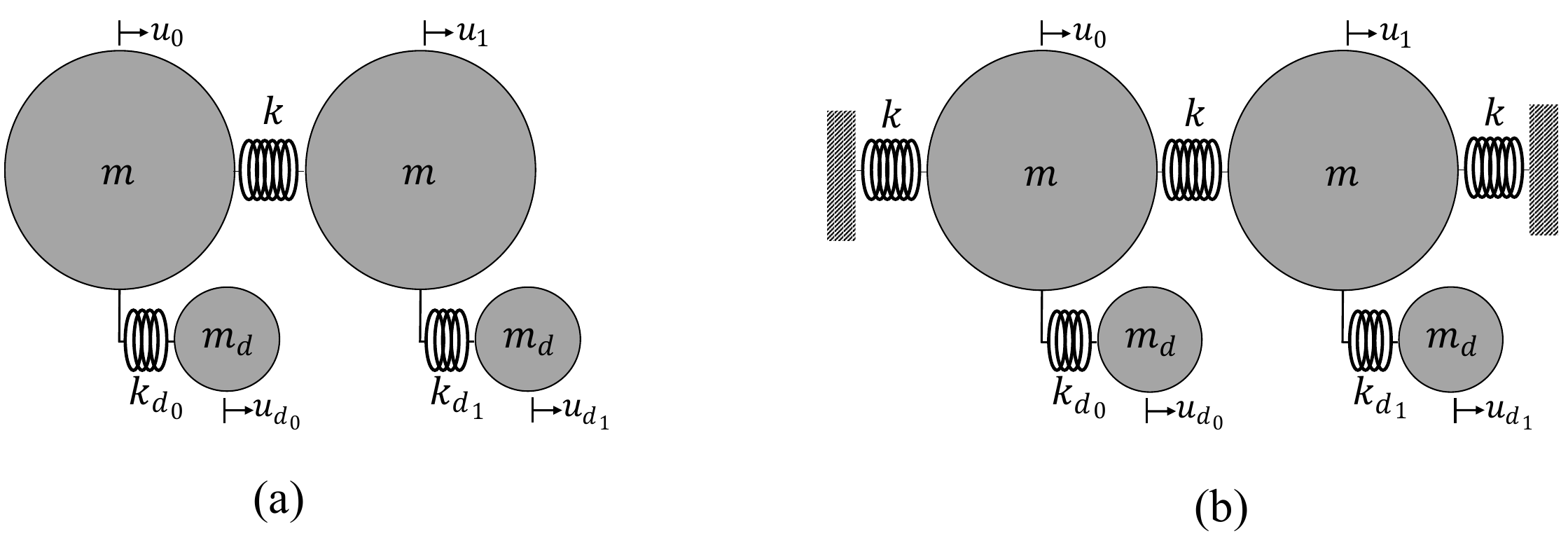}
\caption{The system isolated from its environment. (a) the isolated system with unconstrained/Neumann boundary condition and (b) with fixed/Dirichlet boundary condition.}
\label{naturalfreqs}
\end{figure}

The vibrational eigenfrequencies of the isolated scatterer may be calculated, most conventionally, under free (Fig. \ref{naturalfreqs}a) or fixed (Fig. \ref{naturalfreqs}b) boundary conditions. We will call these sets ${\omega}_N$ (Neumann) and ${\omega}_D$ (Dirichlet) respectively. The aim is to illuminate the relation between $\bar{\omega}_i,\bar{\omega}_i^N,\bar{\omega}_i^D$.
The eigenvalue problems corresponding to the two boundary conditions may be written as:
\begin{equation}
 \label{4dofsystem0}
 \begin{aligned}
 \boldsymbol{H}_{D}\bar{\boldsymbol{u}}=\lambda^D\bar{\boldsymbol{u}}\\
 \boldsymbol{H}_{N}\bar{\boldsymbol{u}}=\lambda^N\bar{\boldsymbol{u}}
 \end{aligned}
\end{equation}
where we use $\boldsymbol{H}_{D}$ and $\boldsymbol{H}_{N}$ to clarify that we are referring to the linear operator (Hamiltonian) of the system isolated from the environment with fixed and free boundary conditions respectively. $\bar{u}$ is the state vector constituting only the degrees of the freedom of the system, $\lambda^D_i$ and $\lambda^N_i$ are the eigenvalues of the isolated system. The corresponding frequency eigenvalues under the two mentioned boundary conditions, $\omega=\sqrt{-\lambda}$, for this problem are given in Table (\ref{table:Eigopenk1}).

\begin{table}[htp]
\begin{tabular}{ |c|c|c|c| } 
 \hline
 ${\omega}_i$  & 0.774661$\pm$ i0.0509381& 1.3706$\pm$ i0.985531 & 2.14715+i0\\ 
 $\omega_i^N$ & 0.742 & 1.338&1.811\\
 $\omega_i^D$ & 0.6067 & 0.806 & 1.557\\
 
 \hline
\end{tabular}
\caption{Eigenvalues of the open system in the first row and the natural frequencies resulted from Neumann and Dirichlet boundary conditions of the central system in second and third rows respectively, for $k=1$ N/m, $k_{d_0}=1.01$ N/m and $k_{d_1}=0.8$ N/m.}
\label{table:Eigopenk1}
\end{table}

 
Comparing $\omega^D_i,\omega^N_i$ to $\bar{\omega}_i$ (Fig. \ref{Transmissionfg}a), we note that there is no clear relation between $\omega^D_i$ and $\bar{\omega}_i$ but there is a close correspondence between $\omega^N_1$ and $\bar{\omega}_1$. It is known that the appearance of the peaks and dips in the scattering spectrum is connected to the existence of discrete poles in the Green’s function of the full scattering problem~\cite{garmon2015bound}. These poles themselves are connected to the eigenvalues of an effective Hamiltonian problem which isolates the system from the environment. The effective problem boundary condition is neither fixed nor free. Such an effective Hamiltonian leads to an eigenvalue problem of the form:
\begin{equation}
    \label{effectiveHamil}
    \boldsymbol{H}_{eff}\bar{\boldsymbol{u}}=\lambda\bar{\boldsymbol{u}}
\end{equation}
where ${\lambda}=-{\omega}^2$. We stress that the eigenvalues resulting from the solution of the above are a special finite set of eigenvalues which correspond to the effective Hamiltonian and which have a special relationship to the scattering of waves. These $\lambda_i$ are related to both of $\lambda_i^D$ and $\lambda_i^N$ through a non-trivial relationship. The eigenvalues, ${\omega}_i$ of the above equation are complex quantities in general. The scattering parameters presented in Fig. (\ref{Transmissionfg}a) are functions of real frequency, however, they have analytical continuation in the complex frequency plane and the eigenvalues ${\omega}_i$ of the effective Hamiltonian are the poles of these analytically continued scattering functions\cite{garmon2015bound}. For this specific problem, the pole locations are provided in Table \ref{table:Eigopenk1} (also shown in Fig. \ref{Transmissionfg}b).

We note that the complex conjugate poles with the real part $0.774661$ lie close to the real line (have small imaginary parts) and their effect is to induce a fast variation of the scattering parameters near $\omega=0.774661$. On the other hand, the complex conjugate poles with the real part $1.3706$ lie away from the real line and their effect is to induce a slow variation of the scattering parameters. We also note that the first pole is close to $\omega_1^N$ but not especially close to $\omega_1^D$. In the subsequent sections, we present a projection operator formalism~\cite{Livsic} for understanding these effect.

\section{projection operator formalism}\label{Feshbachintro}

Projection operators are unitary matrices which project the image of a space onto another space (usually a sub-space) and they are the generalization of the concept of vector projection. Consider an abstract physical system whose state is defined by a vector $\boldsymbol{\psi}$ over some linear vector space $X$ and whose dynamics is characterized by an eigenvalue problem $\boldsymbol{H}\boldsymbol{\psi}=\lambda\boldsymbol{\psi}$. $\boldsymbol{H}$ is an operator in $X$ and is generally known as the Hamiltonian of the problem. We consider two subspaces $U,V$ of $X$ and consider the projection operators $\boldsymbol{P},\boldsymbol{Q}$ which project $\boldsymbol{\psi}$ onto $U,V$ respectively and such that $\boldsymbol{P}+\boldsymbol{Q}=\boldsymbol{I}$ ($\boldsymbol{P}^2=\boldsymbol{P},\boldsymbol{Q}^2=\boldsymbol{Q}$ follow) with $\boldsymbol{I}$ being the identity matrix. This technique allows us to formally extract the effective Hamiltonian, $\boldsymbol{H}_{eff}$, corresponding to the subspace $U$ from the full Hamiltonian $\boldsymbol{H}$. Consider the following two expansions of $\boldsymbol{H}\boldsymbol{\psi}=\lambda\boldsymbol{\psi}$:
\begin{eqnarray}
\boldsymbol{PH}(\boldsymbol{P}+\boldsymbol{Q})\boldsymbol{\psi}=\lambda(\boldsymbol{P}\boldsymbol{\psi})\\
\boldsymbol{QH}(\boldsymbol{P}+\boldsymbol{Q})\boldsymbol{\psi}=\lambda(\boldsymbol{Q}\boldsymbol{\psi})
\end{eqnarray}
Using $\boldsymbol{P}^2=\boldsymbol{P},\boldsymbol{Q}^2=\boldsymbol{Q}$, we have:
\begin{eqnarray}
\label{fesh1}
\boldsymbol{PHP}(\boldsymbol{P}\boldsymbol{\psi})+\boldsymbol{PHQ}(\boldsymbol{Q}\boldsymbol{\psi})=\lambda(\boldsymbol{P}\boldsymbol{\psi})\\
\label{fesh2}
\boldsymbol{QHP}(\boldsymbol{P}\boldsymbol{\psi})+\boldsymbol{QHQ}(\boldsymbol{Q}\boldsymbol{\psi})=\lambda(\boldsymbol{Q}\boldsymbol{\psi}),
\end{eqnarray}
resulting in:
\begin{equation}
    \boldsymbol{Q}\boldsymbol{\psi}=\frac{1}{\lambda-\boldsymbol{QHP}}(\boldsymbol{P}\boldsymbol{\psi}).
\end{equation}
Substituting the above into the Eq. (\ref{fesh1}) we obtain:
\begin{equation}
    \left(\boldsymbol{PHP}+\boldsymbol{PHQ}\frac{1}{\lambda-\boldsymbol{QHQ}}\boldsymbol{QHP}\right)(\boldsymbol{P}\boldsymbol{\psi})=\lambda(\boldsymbol{P}\boldsymbol{\psi})
\end{equation}
The above is an eigenvalue equation whose eigenvectors are $\boldsymbol{P}\boldsymbol{\psi}$ which belong to the subspace $U$. The operator which acts on $\boldsymbol{P}\boldsymbol{\psi}$ on the left hand side of the equation can be identified as the effective Hamiltonian. Thus, $\boldsymbol{H}_{eff}=\boldsymbol{PHP}+\boldsymbol{\Sigma}(\lambda)$ where $\boldsymbol{\Sigma}(\lambda)$ is called ``self-energy" in quantum mechanics\cite{sasada2008calculation} and is a measure of the coupling of the system with its environment. $\boldsymbol{\Sigma}(\lambda)$ is given by:
\begin{equation}
    \boldsymbol{\Sigma}(\lambda):=\boldsymbol{PHQ}\frac{1}{\lambda-\boldsymbol{QHQ}}\boldsymbol{QHP}
    \label{selfenergy}
\end{equation}
$\boldsymbol{H}_{eff}$ can thus be calculated by substituting Eq. (\ref{selfenergy}) and the expression for $\boldsymbol{PHP}$ into $\boldsymbol{H}_{eff}=\boldsymbol{PHP}+\boldsymbol{\Sigma}(\lambda)$.

\section{1D discrete problem}\label{section5}
\begin{figure}[htp]
\centering
\includegraphics[scale=.45]{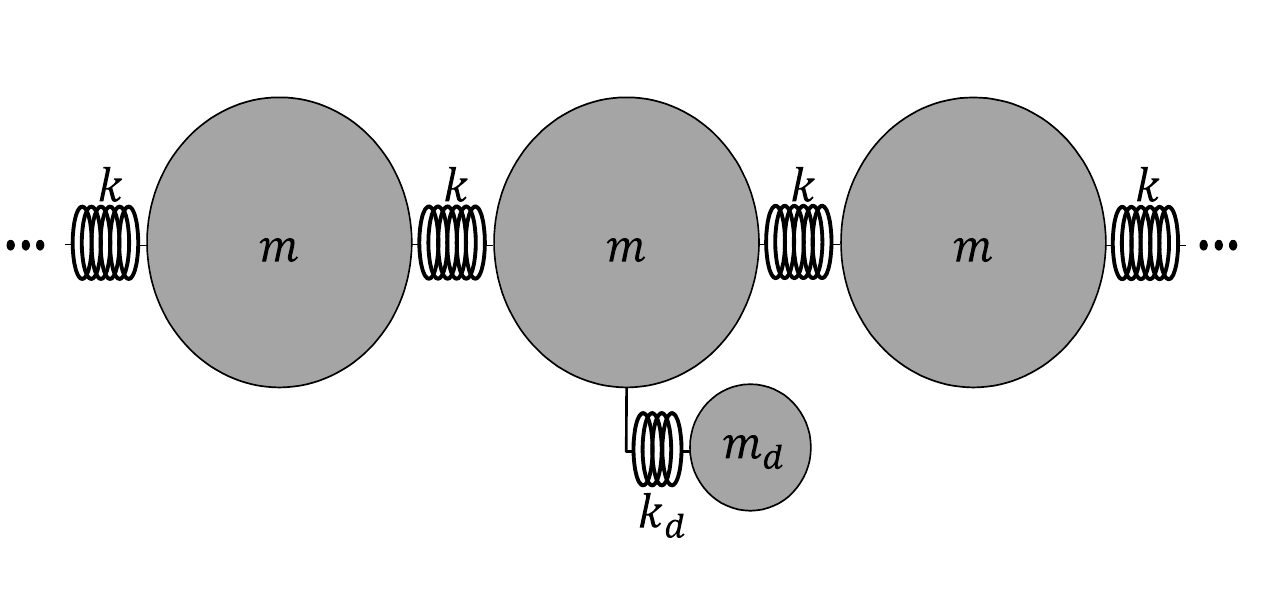}
\caption{1D spring mass model. Mass $m_d$ and and two semi infinite chains are connected to the site $0$}\label{1D_model}
\end{figure}

To see an explicit application of the projection operator formalism, consider the case of a simple 1D discrete system as shown in Fig.(\ref{1D_model}). This problem consists of an infinite chain of masses, $m$, connected to their neighbors through a spring with spring constant $k$. These masses are indexed with the integer $n=-\infty,...-1,0,1...\infty$. A single mass (at site 0) has a further connection to another mass, $m_d$, through a spring with spring constant $k_d$ and it acts as the scatterer. We first derive the effective Hamiltonian for this problem directly and then derive it using the projection operators. 

\subsection{Direct method using Siegert boundary conditions}

We can write the equation of motion for this system as follows:

\begin{equation}
\label{EoM}
\left\{
\begin{array}{lcl}
  m\Ddot{u_n}=k(u_{n-1}+u_{n+1}-2u_n)  & \mbox{for} & n\neq 0\\
  m\Ddot{u_0}=k(u_{-1}+u_{1}-2u_0)+k_d(u_d-u_0) & \mbox{for} & n=0\\
  m_d\Ddot{u_d}=k_d(u_0-u_d)&\mbox{for}& m_d
\end{array}\right.
\end{equation}


The equation of motion (\ref{EoM}) can be written in the abstract form $\ddot{\boldsymbol{u}}=H\boldsymbol{u}$ where $H$ is the Hamiltonian of the system. Assuming $m=m_d=1$, we can write this Hamiltonian projected on the orthonormal basis {\color{red}(in matrix form)}, as:



\begin{equation}
\label{hamiltonian}
\boldsymbol{H}=
    \begin{pmatrix}
    \ddots&\ddots&&&&\\
    &k&-2k&k&&\\
    &&k&-2k-k_d&k_d&k&\\
    &&0&k_d&-k_d&0&\\
    &&&k&0&-2k&k\\
    &&&&&\ddots&\ddots
    \end{pmatrix}
\end{equation}

Instead of solving the second order time dependent problem, we can consider the eigenvalue form of the problem given by $\boldsymbol{H}\boldsymbol{u}=\lambda\boldsymbol{u}$ with $\lambda=-\omega^2$  where $\boldsymbol{u}=\left[...,u_{-1},u_{0},u_{d},u_{1},...\right]$ and $\omega$ is the angular frequency in radians per second. In the rest of this paper, we will consider various angular frequencies and their units will always be radians per second, and the units will, in general, not be explicitly mentioned. We will consider wave scattering problems in this system and such problems are characterized by reflected and transmitted waves in the following form:
\begin{equation}
    u_n=\left\{\begin{array}{lcl}
         Ae^{inqL}+Be^{-inqL}& \mbox{for} & n\leq-1 \\
         Ce^{inqL}+De^{-inqL}& \mbox{for} & n\geq1 
    \end{array}
    \right.
\end{equation}

where $q$ is the wavenumber and $L$ is the distance between the adjacent masses in the chain. Here, $Ae^{inqL},De^{-inqL}$ represent waves which are incident towards the central system (incoming waves) from the left and the right, respectively, and $Be^{-inqL},Ce^{inqL}$ represent waves which are traveling away from the central system (outgoing waves). If we are interested in extracting from the above, only those eigenvalues which correspond to the effective Hamiltonian, then we are only interested in analyzing the resonant states of the scattering problem (corresponding to the poles of the scattering coefficients). It is known that the resonant states can be extracted if one considers only the outgoing waves in the problem (Siegert boundary conditions\cite{siegert1939derivation,hatano2013equivalence}):

\begin{equation}
    u_n=\left\{\begin{array}{lcl}
         Be^{-inqL}& \mbox{for} & n\leq-1 \\
         Ce^{inqL}& \mbox{for} & n\geq1 
    \end{array}
    \right.
\end{equation}

For sites $n\geq2$, we can write the Hamiltonian in Eq.(\ref{hamiltonian}):
\begin{equation}
\omega^2Ce^{inqL}=kCe^{inqL}(2-e^{-iqL}-e^{iqL})=2kC(1-\cos{(qL)})
\end{equation}
which gives rise to the dispersion relation $\omega^2=2k(1-\cos{(qL)})$. Dispersion relation for $n\leq2$ is exactly the same in this case. These dispersion relations connect the wavenumber and frequency of the wave scattered in the environment. From the continuity of displacement at the site $n=0$ we conclude that $B=C=u_0$. Now we write the Hamiltonian for the sites $0$ and $d$:
\begin{equation}
\label{site_0}
   \left\{\begin{array}{lcl}
        - \omega^2u_0+k(2u_0-u_{-1}-u_{1})+k_d(u_0-u_d)=0 \\
         -\omega^2u_d+k_d(u_d-u_0)=0
    \end{array}
    \right.
\end{equation}
We would like to extract the degrees of freedom of the system, ($u_0,u_d$), from the above. This can be done by noting that $u_{\pm1}=e^{iqL}u_0$. Substituting this in Eq.(\ref{site_0}), we have:
\begin{equation}
\left\{\begin{array}{lcl}
         -k(2u_0-e^{iqL}u_0-e^{iqL}u_0)-k_d(u_0-u_d)=-\omega^2u_0\\
         -k_d(u_d-u_0)=-\omega^2u_d
    \end{array}
    \right.
\end{equation}
After some rearrangement:
\begin{equation}
    \begin{pmatrix}
    -2k(1-e^{iqL})-k_d & k_d\\
    k_d &- k_d
    \end{pmatrix}\begin{pmatrix}
    u_0\\
    u_d
    \end{pmatrix}
    = -\omega^2\begin{pmatrix}
    u_0\\
    u_d
    \end{pmatrix}
    \label{eig_prob}
\end{equation}
The above equation is of the form $\boldsymbol{H}_{eff}\bar{\boldsymbol{u}}={\lambda}\bar{\boldsymbol{u}}$ where ${\lambda}=-{\omega}^2$ and the effective Hamiltonian is explicitly given by:
\begin{equation}
\label{H_eff}
    \boldsymbol{H}_{eff}=\begin{pmatrix}
    -2k(1-e^{iqL})-k_d & k_d\\
    k_d &- k_d
    \end{pmatrix}
\end{equation}
Thus, we have been able to directly extract the effective Hamiltonian of the system by employing only the outgoing waves (Siegert boundary conditions).

\subsection{Effective Hamiltonian using Projection Operators}
Now we derive the effective Hamiltonian using the projection operators described in Sec. (\ref{Feshbachintro}). We first define appropriate projection operators $\boldsymbol{P}$ and $\boldsymbol{Q}$ where $\boldsymbol{P}$ is expected to isolate the degrees of freedom of the system, and can be appropriately understood as an infinite dimensional square matrix. $\boldsymbol{Q}$ is the dual of $\boldsymbol{P}$. Explicitly they are given by:

\begin{eqnarray}
\nonumber \boldsymbol{P}:=\begin{pmatrix}
0&&&&&&\\
&\ddots&&&&&\\
&&1&&&&\\
&&&1&&&\\
&&&&\ddots&&\\
&&&&&&0
\end{pmatrix},\quad
\boldsymbol{Q}=\boldsymbol{I}-\boldsymbol{P}=\begin{pmatrix}
1&&&&&&\\
&\ddots&&&&&\\
&&0&&&&\\
&&&0&&&\\
&&&&\ddots&&\\
&&&&&&1
\end{pmatrix}
\end{eqnarray}

The non-zero diagonal terms in $\boldsymbol{P}$ (and the zero diagonal terms in $\boldsymbol{Q}$) exist at locations corresponding to the degrees of freedom of the central system. The effective Hamiltonian from the projection operator formalism  is:
\begin{equation}
    \boldsymbol{H}_{eff}=\boldsymbol{PHP}+\boldsymbol{PHQ}\frac{1}{{\lambda}-\boldsymbol{QHQ}}\boldsymbol{QHP}
\end{equation}
The effective Hamiltonian is of the form $\boldsymbol{PHP}+\boldsymbol{\Sigma}({\lambda})$ where: (see\cite{sasada2011resonant} for details):
\begin{eqnarray}
\boldsymbol{PHP}=\begin{pmatrix}
-2k-k_d & k_d\\k_d&-k_d
\end{pmatrix};
\qquad
\boldsymbol{\Sigma}=\begin{pmatrix}
2ke^{iqL}& 0\\0 & 0
\end{pmatrix}
\end{eqnarray}

Thus the effective Hamiltonian comprises of two parts. The first part ($\boldsymbol{PHP}$) is hermitian and it is precisely the Hamiltonian of the system if it were isolated from the environment by applying fixed boundary conditions at its ends. The second part $\boldsymbol{\Sigma}$ is non-hermitian and it emerges as a correction to $\boldsymbol{PHP}$ and accounts for the interaction between the system and the environment. The non-hermiticity of $\boldsymbol{\Sigma}$ is the source of apparent damping as one notices the energy in the system escaping into the environment. Combining, we get:
\begin{equation}
\boldsymbol{H}_{eff}=\boldsymbol{PHP}+\boldsymbol{\Sigma}=\begin{pmatrix}
-2k-k_d & k_d\\k_d&-k_d
\end{pmatrix}+\begin{pmatrix}
2ke^{iqL}& 0\\0 & 0
\end{pmatrix}=\begin{pmatrix}
    -2k(1-e^{iqL})-k_d & k_d\\
    k_d &- k_d\end{pmatrix}
    \label{EffecHamilexp}
\end{equation}
which is precisely the expression calculated earlier from the direct method (Eq. \ref{H_eff}). However, going through the route of the  projection operator formalism has exposed the deep physical insight that the effective Hamiltonian is a correction over the Hamiltonian of the isolated system with fixed boundary conditions - an insight that was hidden in the direct method.

At this point, we note that the entire physics of the scattering problem in the open system is encapsulated in the finite dimensional effective Hamiltonian $\boldsymbol{H}_{eff}$. We would like to find the eigenvalues of this effective Hamiltonian and, finally, we would like to find the Green's function of the same as it would allow us to connect these eigenvalues to the peaks and dips of the scattering spectrum. However, $\boldsymbol{H}_{eff}$ itself is not a simple linear operator due to its dependence on the wavenumber $q$, which makes it dependent on the frequency $\omega$ through the dispersion relation. Thus, finding the eigenvalues of $\boldsymbol{H}_{eff}$ is not a straightforward exercise. $\boldsymbol{H}_{eff}$ can be written as a quadratic eigenvalue problem. To do so, we take $\beta=e^{iqL}$ and note that $\boldsymbol{H}_{eff} \bar{\boldsymbol{u}}={\lambda}\bar{\boldsymbol{u}}$ is equivalent to the following:
\begin{equation}
\label{quadratic}
    \begin{pmatrix}
    -2k(1-\beta)-k_d & k_d\\
    k_d &- k_d
    \end{pmatrix}\begin{pmatrix}
    u_0\\
    u_d
    \end{pmatrix}
    = -\left [ 2k-k(\beta+\frac{1}{\beta})\right ]\begin{pmatrix}
    u_0\\
    u_d
    \end{pmatrix}    
\end{equation}
The derivation of the right hand side exploits the dispersion relation. We identify the above as a quadratic eigenvalue problem in $\beta$ through the following rearrangement:
\begin{equation}
    \left[ \beta^2\underbrace{\begin{pmatrix}
    k& 0\\
    0 & -k
    \end{pmatrix}}_M+\beta\underbrace{\begin{pmatrix}
    -k_d & k_d\\
    k_d &-k_d+2k
    \end{pmatrix}}_Y+\underbrace{\begin{pmatrix}
    -k & 0\\
    0 & -k
    \end{pmatrix}}_K\right]\begin{pmatrix}
    u_0\\
    u_d
    \end{pmatrix}  =0 
\end{equation}

We refer to this form of the effective Hamiltonian as $\boldsymbol{Q}(\beta)\bar{\boldsymbol{u}}=0$. Since in this problem, $\bar{u}$ has 2 elements, we note that there are 4 eigenvalues of the quadratic eigenvalue problem $\boldsymbol{Q}(\beta)\bar{\boldsymbol{u}}=0$. We denote these eigenvalues as $\beta_i,i=1,...4$. Each $\beta_i$ corresponds to a certain ${\omega}_i$ through the dispersion relation. In general, if there are $n$ degrees of freedom in the system, then there are $2n$ eigenvalues of the corresponding effective Hamiltonian. These eigenvalues are either real or they come in complex conjugate pairs as long as the matrices $\boldsymbol{M},\boldsymbol{Y},\boldsymbol{K}$ are all hermitian, which is the case in the absence of any explicit damping elements. There are standard ways of solving the quadratic eigenvalue problem and one method is to transform the above into a linear matrix pencil\cite{tisseur2001quadratic}. We will not go into the details of its implementation as they are standard. 

One can now consider a scattering problem where some wave with wave number and frequency $q,\omega$ (or equivalently $q$ and $\lambda=-\omega^2$) incident on the system gets scattered into reflected and transmitted waves. In this case, $q,\omega$ are related through the dispersion relation in the environment but they need not have any relation to $\beta_i,{q}_i,{\omega}_i$ which are the specific eigenvalues of the effective Hamiltonian. In this case, the system response is fully characterized by a forced problem of the form $(\lambda-\boldsymbol{H}_{eff})\bar{\boldsymbol{u}}=0$. The solutions to this forced problem should provide the scattering coefficients in the environment. To illuminate the precise relationship, we can first calculate the Green's function of this problem. The Green's function of the problem can be found by taking the inverse of the associated operator \cite{hatano2013equivalence}:
\begin{equation}
\label{Greensfraction}
    \boldsymbol{G}(\omega)=\frac{1}{\lambda-\boldsymbol{H}_{eff}}
\end{equation}

We now note that the following relation exists from (\ref{quadratic}):

\begin{equation}
\label{Green}
    \lambda-\boldsymbol{H}_{eff}=-\frac{\boldsymbol{Q}(\beta)}{\beta}
\end{equation}

We now expand the Green's function in its modal expansion \cite{tisseur2001quadratic}:

\begin{equation}
    \boldsymbol{G}(\omega)=-\frac{\beta}{\boldsymbol{Q}(\beta)}=\sum^{4}_{i=1} \boldsymbol{x}_i\frac{\beta\beta_i}{\beta-\beta_i}\boldsymbol{y}_i^T
    \label{Greens}
\end{equation}

where $\boldsymbol{x}_i,\boldsymbol{y}_i$ are the right and left eigenvectors of the quadratic eigenvalue problem. In the present problem, $\boldsymbol{x}_i\boldsymbol{y}_i^T$ is a $2\times 2$ square matrix (since the dimension of $\bar{\boldsymbol{u}}$ is 2) and, therefore, the Green's function, $\boldsymbol{G}(\omega)$, is also a $2\times 2$ matrix, more appropriately written as $G_{ij};i,j=1,2$. The above expression makes it clear the the dynamic response of the system, as encapsulated in the Green's function $G_{ij}(\omega)$, has poles at the eigenvalues of the effective Hamiltonian. Therefore, if one were to excite the system with a frequency $\omega$ which lines up with a pole $\beta_i$, one would expect to see resonance phenomenon in the scattering parameters. 

\begin{figure}[htp]
\centering
\includegraphics[scale=.45]{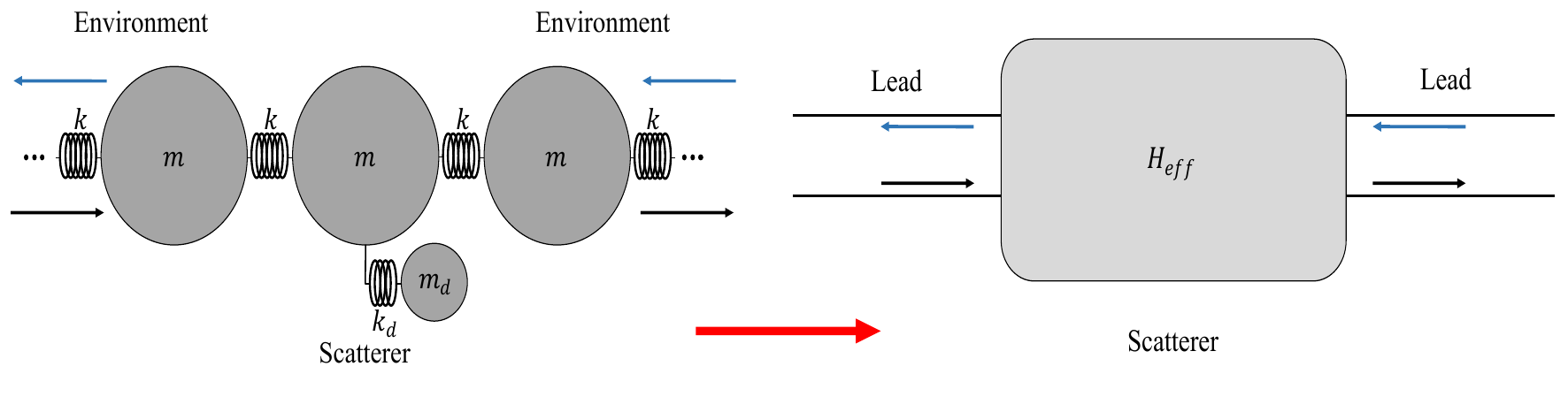}
\caption{Scattering from a central system in two semi-infinite chains on the left and scattering in two leads connected to a central system on the right.  }\label{leadandenv}
\end{figure}

The entire system can now be considered as a scatterer connected to two leads as shown in Fig. (\ref{leadandenv}). Each lead admits left and right traveling waves with wave characteristics corresponding to the waves in the environment. Since the environment on the two sides of the system is the same in this problem and since there is only one wavenumber solution $q$ in the environment at each frequency $\omega$, each lead admits a single left and a single right traveling wave at a given frequency $\omega$. These waves are characterized by wavenumbers $q,-q$ with $q$ being related to $\omega$ through the dispersion relation of the leads (environment). The scattering matrix $\boldsymbol{S}$, in this case, is a $2\times 2$ matrix more properly written as $S_{ij};i,j=1,2$. $S_{21}$, for example, represents the transmission amplitude and $S_{11}$ represents the reflection amplitude when the incident wave is coming from the left. The Fisher-Lee relation\cite{fisher1981relation} connects the components of the scattering matrix with the components of the Green's function matrix of the system through a proportional relationship $S_{ij}\propto G_{ij}$ showing clearly that resonance in scattering is expected in the vicinity of the eigenvalues of the effective Hamiltonian. The relation can be explicitly calculated for the present discrete problem through a process similar to the direct method of finding effective Hamiltonian. We consider incident (from the left) and scattered waves in the system:
\begin{equation}
\label{reftrans1}
    u_n=\left\{\begin{array}{lcl}
         Ae^{inqL}+Be^{-inqL}& \mbox{for} & n\leq-1 \\
         Ce^{inqL}& \mbox{for} & n\geq1 
    \end{array}
    \right.
\end{equation}
Continuity condition at $n=0$, gives $A+B=C=u_0$, which leads to:
\begin{equation}
    \label{reftrans2}
    \begin{aligned}
        u_{-1}=Ae^{-iql}+\left(u_0-A\right)e^{iql}\\
        u_1=Ce^{iql}
    \end{aligned}
\end{equation}
By substituting the expressions for $u_{-1}$ and $u_1$ from (\ref{reftrans1}) into Eq. (\ref{site_0}), we can write after some manipulations:
\begin{equation}
    \label{reftrans3}
    \left(\lambda-\boldsymbol{H}_{eff}\right)
    \begin{pmatrix}
    u_0\\
    u_d
    \end{pmatrix}=-2Aki\sin ql
    \begin{pmatrix}
    1\\0
    \end{pmatrix}
\end{equation}
Here, by paying attention to Eq. (\ref{Greensfraction}), for the $2\times 2$ Green's matrix we can write:
\begin{equation}
    \label{reftrans4}
    \begin{pmatrix}
    u_0\\
    u_d
    \end{pmatrix}=-2Aki\sin ql
    \begin{pmatrix}
    G_{11}\\G_{21}
    \end{pmatrix}
\end{equation}
The scattering matrix coefficients are $S_{11}=B/A,S_{21}=C/A$ which can now be represented as:
\begin{equation}
    \label{reftrans5}
    \begin{aligned}
        S_{11}=-2ik\sin ql G_{11}-1\\
        S_{21}=-2ik\sin ql G_{11}
    \end{aligned}
\end{equation}

The above shows the scattering resonances are expected in the vicinity of the eigenvalues of the effective Hamiltonian.

\subsection{Fixed and Free boundary conditions}

We first note that in the effective Hamiltonian equation $\boldsymbol{H}_{eff}=\boldsymbol{PHP}+\boldsymbol{\Sigma}({\lambda})$, $\boldsymbol{PHP}$ is the Hamiltonian of the isolated system under fixed boundary conditions. Thus $\boldsymbol{H}_{eff}=\boldsymbol{H}_D+\boldsymbol{\Sigma}$. This equation can be further reorganized: 
\begin{equation}
\boldsymbol{H}_{eff}=\begin{pmatrix}
-k_d & k_d\\k_d&-k_d
\end{pmatrix}-\begin{pmatrix}
2k & 0\\0&0
\end{pmatrix}+e^{iqL}\begin{pmatrix}
2k& 0\\0 & 0
\end{pmatrix}
    \label{Neumann1}
\end{equation}
The first matrix in the above is the effective Hamiltonian of the isolated scatterer with free boundary conditions ($\boldsymbol{H}_N$). Thus, we have:
\begin{equation}
\boldsymbol{H}_{eff}=\boldsymbol{H}_N+(e^{iqL}-1)\tilde{\boldsymbol{K}}   
\label{Neumann3}
\end{equation}
where, $\tilde{\boldsymbol{K}}$ is a diagonal matrix with two non-zero elements equal to $k$ on the diagonal at $(0,0)$ and $(N,N)$. Eq. (\ref{Neumann3}) can be used to arrive at an interesting result relating to the location of the scattering resonances. Since these locations are directly dependent upon the eigenvalues of $\boldsymbol{H}_{eff}$, whenever $\boldsymbol{H}_{eff}\approx \boldsymbol{H}_N$, these resonances can be determined by the eigenvalues of the scatterer under free boundary conditions. This happens when $e^{iql} \approx 1$ or in the low frequency regime. Therefore, if there are eigenvalues of $\boldsymbol{H}_N$ in the low frequency regime then one can expect scattering resonances close to them. In fact, considering the Euclidean norm of second matrix in Eq. (\ref{Neumann3}), $(1-e^{iql})\tilde{\boldsymbol{K}}$, as $2k(1-\cos{ql})$, which according to the dispersion relation is $\omega^2$, at small eigenvalues the eigenvalue problem of the open system can be treated as a problem of perturbed isolated system with Neumann boundary conditions.

\section{Numerical Results and Discussion}\label{NR}
\begin{figure}[htp]
\centering
\includegraphics[scale=.45]{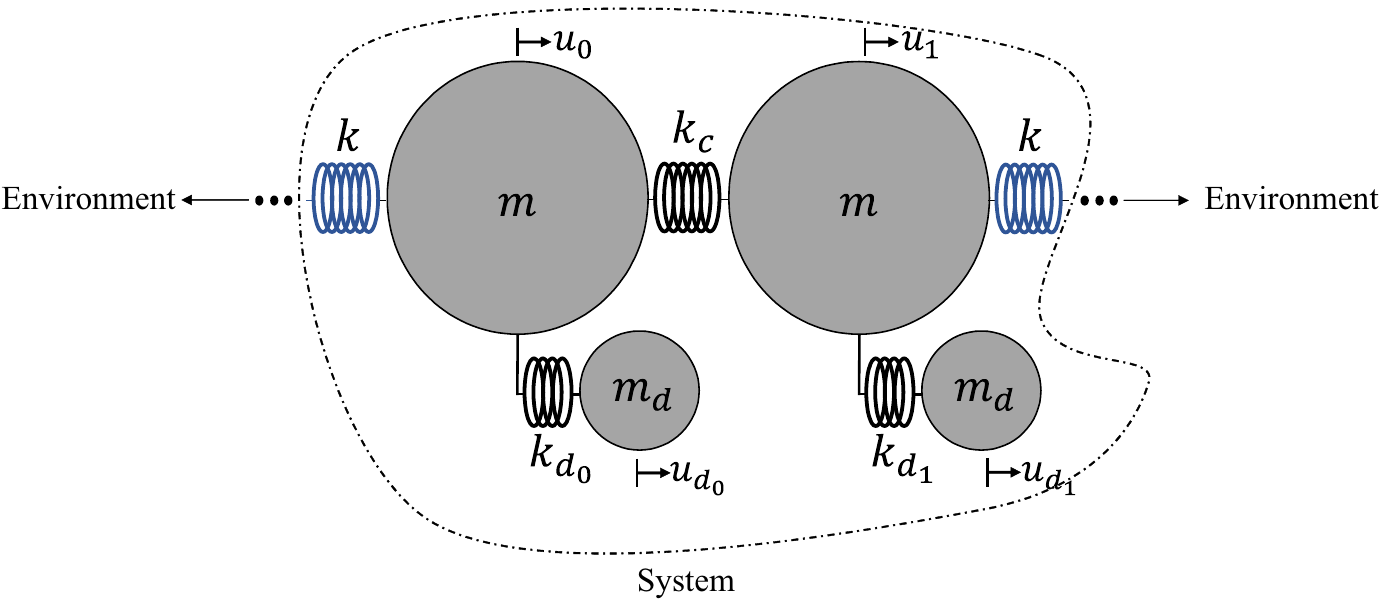}
\caption{A mechanical open system with two resonators in touch with the environment with two symmetric chains.}\label{system1}
\end{figure}

We now present a set of numerical examples which illuminate some of the salient points from the last few sections. The examples are based upon the general setup shown in Fig. (\ref{system1}) which is also similar to the numerical example in section \ref{section2}. The main difference here is the inclusion of a potentially different spring in the interior of the system, now denoted by $k_c$. 

In section \ref{section2}, we considered the case where all the spring constants in the main chain were the same with the value of $k=1$ N/m, and the spring constants in the left and right resonators were $k_{d_0}=1.01$ N/m and $k_{d_{1}}=0.8$ N/m, respectively. The aim here is to manipulate the properties of the system and the environment to affect certain desired changes in the scattering spectrum. For instance, we noted that the magnitude of the imaginary parts of $\omega_i$ has an impact on the sharpness and locations of the scattering peaks~\cite{hatano2008some}. As an example, a choice of $k=0.8$ N/m, $k_c=2$ N/m, $k_{d_0}=2$ N/m and $k_{d_1}=0.8$ N/m results in all eigenvalues, $\omega_i$, having the following imaginary parts:

 

\begin{table}[h!]
\begin{tabular}{ |c|c|c|c| } 
 \hline
 ${\omega}_i$ & 0.902$\pm$ i0.0433 & 1.854+i0 & 2.643+i0\\
  $\omega_i^N$ & 0.887 & 1.622 &2.486\\
 
 \hline
\end{tabular}
\caption{Eigenvalues of the open system in the first row and the natural frequencies of the isolated system, under Neumann boundary conditions, in the second row, for $k=0.8$ N/m, $k_c=2$ N/m,  $k_{d_0}=2$ N/m and $k_{d_1}=0.8$ N/m.}
\label{table:Eigopenk0.8}
\end{table}

 

\begin{figure}[h!]
\centering
\includegraphics[scale=0.182]{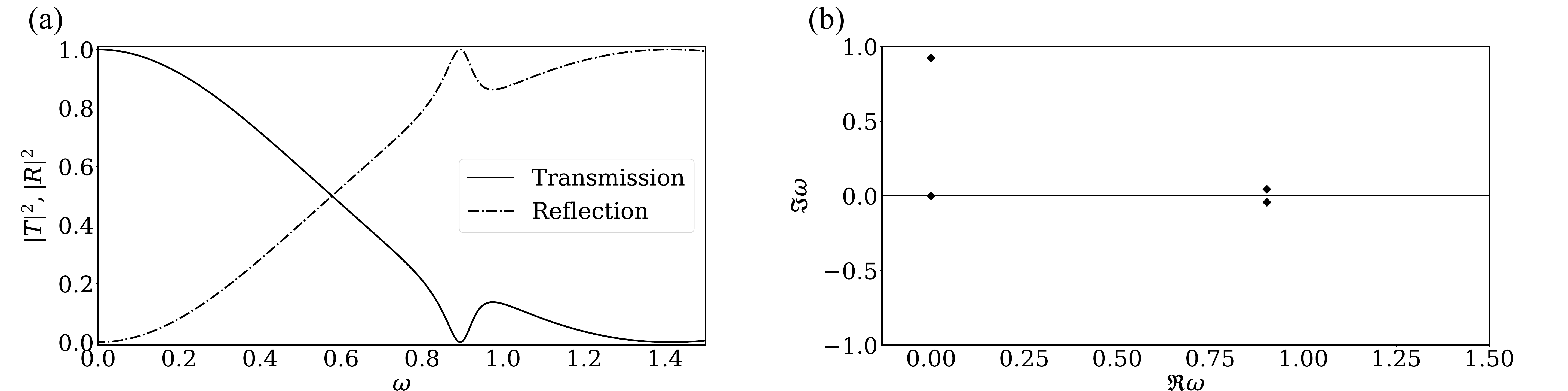}
\caption{  For $k=0.8$ N/m, $k_c=2$ N/m, $k_{d_0}=2$ N/m and $k_{d_1}=0.8$ N/m (a) the transmission/reflection amplitude vs. angular frequency is shown and (b) is representing the imaginary vs. real parts of discrete poles.}\label{Transmissionfgk0.8}
\end{figure}

We note that the effective Hamiltonian results in eight eigenvalues since $\bar{u}$ has a dimension of four in this problem, however, Table \ref{table:Eigopenk0.8} lists only four of them. Fig. (\ref{Transmissionfgk0.8}) shows the transmission and reflection spectra when the system has the above properties and it shows a prominent peak/dip in the spectrum at a frequency of $\bar{\omega}_1=0.894$. From Table (\ref{table:Eigopenk0.8}), we note that there is a corresponding $\omega_i=0.902\pm 0.0433$ near this peak. The small imaginary part of this particular pole results in a small width (sharp peak) of the associated scattering resonance~\cite{Rotter_2009}. The precise location of the peak/dip, $\bar{\omega}=0.894$ rad/sec, exhibits a shift from the location of the closest vibrational eigenvalue of the system under Neumann boundary conditions as, $\omega_1^N=0.887$ rad and this shift is mediated through both the system interacting with its environment and also the meromorphic expression in Eq.(\ref{Greens}). The former effect not only makes the eigenvalues complex but also shifts their real parts compared to the eigenvalues of an isolated system ($\omega_i^N\rightarrow\omega_i$). This effect is mediated through the effective Hamiltonian relationship given in Eq.  (\ref{Neumann3}). The latter effect results in a further shift $\omega_i\rightarrow\bar{\omega}_i$. The first shift can be interpreted as the effect of the environment on the eigenvalues of the isolated system~\cite{LinearPerturbation}, while the second shift stems from the point that a transmission/reflection function is a function of the Green's function, which itself is a meromorphic function of all the discrete poles of the effective Hamiltonian. 

\subsection{Sharp peaks and broad peaks}

\begin{figure}[htp]
\centering
\includegraphics[scale=0.18]{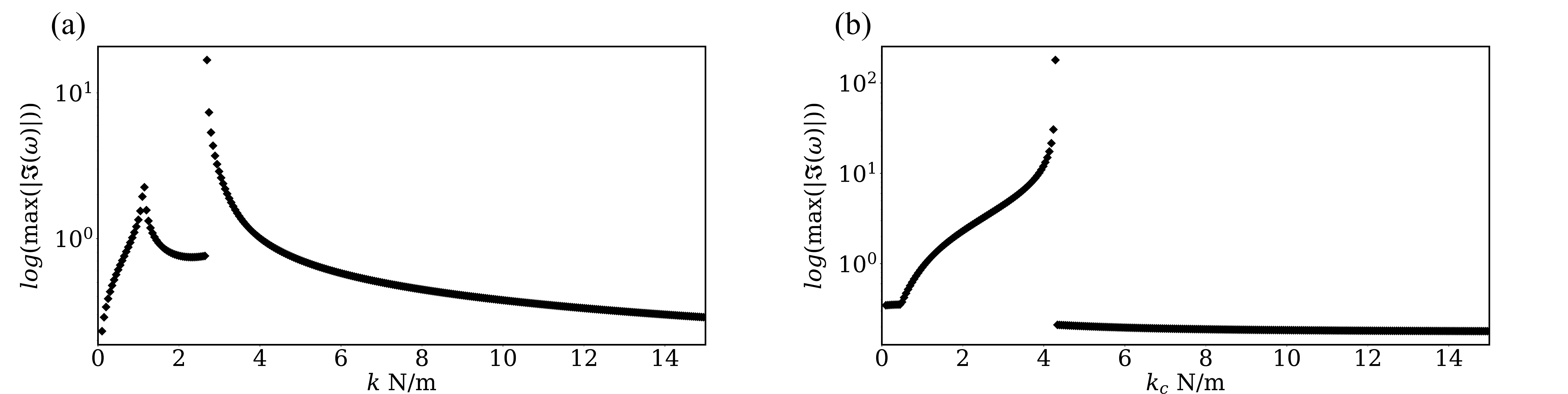}
\caption{ For a 4-DOF scatterer with $k_{d_0}=2$ N/m and $k_{d_1}=0.8$ N/m, (a) is showing the change in maximum of the imaginary part of eigenvalues vs. the stiffness in the environment, $k$, with $k_c=0.5$ N/m and (b) is showing the change in maximum of the imaginary part of eigenvalues vs. the stiffness in the main chain of the scatterer, $k_c$, with $k=15$ N/m.}\label{maximgandmaxdeltrealvskforkc3}
\end{figure}

We can now try to manipulate the properties of the problem to achieve some desired properties in the scattering spectrum. For instance, we would like to achieve a scattering spectrum with sharp peaks/dips as such sharp transitions may be useful for incorporating as sensing mechanisms\cite{wang2021angledependent}. There are several ways of doing this but we will focus on two mechanisms - varying $k$ and $k_c$. Variation of $k_c$ strictly affects the system since it only appears in the $\boldsymbol{PHP}$ part of the effective Hamiltonian, whereas varying $k$ affects the coupling with the environment as well. For the problem under consideration, Fig. (\ref{maximgandmaxdeltrealvskforkc3}a) for a system with $k_c=0.5$ N/m, shows in logarithmic scale the change of maximum imaginary part among the eigenvalues of the open system vs. the change of spring stiffness in the environment, $k$. Also for $k=15$ N/m,  Fig. (\ref{maximgandmaxdeltrealvskforkc3}b) shows the change of maximum imaginary part among the eigenvalues of the open system vs. the change of spring stiffness in the the main chain of the scatterer, $k_c$. These figures suggests some mechanisms with which to affect the sharpness of the scattering transitions.

\begin{table}[htp]
\begin{tabular}{ |c|c|c|c|c|c|c| } 
 \hline
\multicolumn{4}{|c|}{System 1} &\multicolumn{3}{|c|}{System 2}\\
 \hline\hline
 ${\omega}_i$  & 0.898$\pm$ i0.046& 1.384$\pm$ i0.148  & 8.445+i0& 0.734$\pm$i0.196&  1.077$\pm$i0.762 & 2.9057+i0\\ 
  $\omega_i^N$ & 0.979 & 1.761 & 4.641& 0.626 & 1.376 & 2.077 \\
  \hline
\end{tabular}
\caption{The Eigenvalues for two open systems and for the systems under free boundary conditions. The first three columns are for a system with $k=15$ N/m, $k_c=10$ N/m,  $k_{d_0}=2$ N/m and $k_{d_1}=0.8$ N/m and the second three columns are for a system with $k=2 $ N/m, $k_c=0.5$ N/m,  $k_{d_0}=2$ N/m and $k_{d_1}=0.8$ N/m.}
\label{table:Eigopenk15}
\end{table}



 

\begin{figure}[htp]
\centering
\includegraphics[scale=0.18]{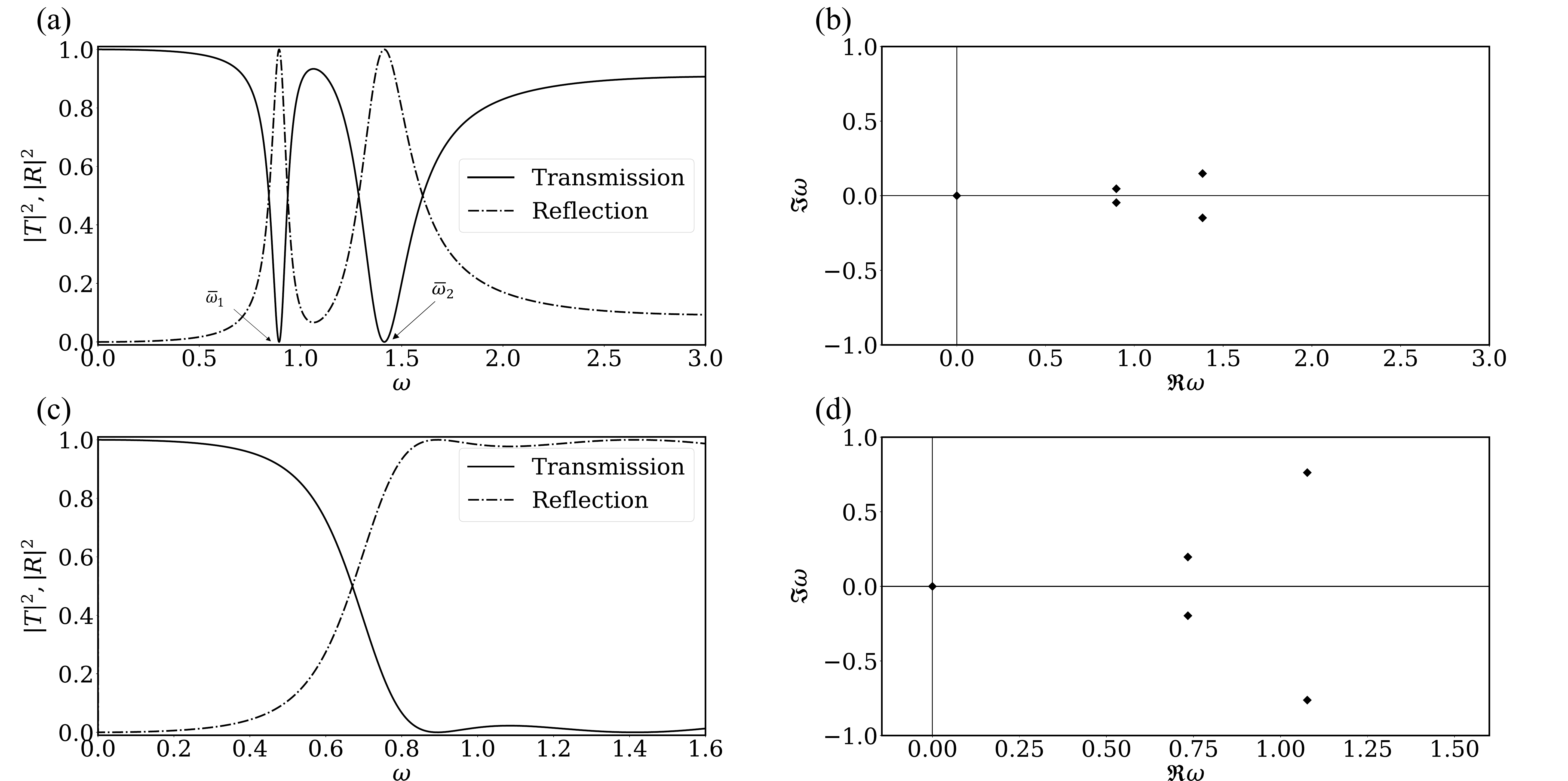}
\caption{ For $k=15$ N/m, $k_c=10$ N/m, $k_{d_0}=2$ N/m and $k_{d_1}=0.8$ N/m (a) is showing the transmission/reflection vs. angular frequency and (b) is representing the imaginary vs. the real parts of discrete poles. For $k=2$ N/m, $k_c=0.5$ N/m, $k_{d_0}=2 $ N/m and $k_{d_1}=0.8$ N/m (c) is showing the transmission/reflection vs. angular frequency and (d) is representing the imaginary vs. the real parts of discrete poles.}\label{Transmissionfgk8}
\end{figure}

As an example, $k=15$ N/m at $k_c=10$ N/m forces the $\omega_i$ to have small imaginary parts and they are plotted in Fig. (\ref{Transmissionfgk8}b) and shown in Table \ref{table:Eigopenk15}. In Fig. (\ref{Transmissionfgk8}a), we show that the peaks and valleys corresponding to the scattering spectrum in this problem are sharp -- in fact they are sharper for $\bar{\omega}_1$ than $\bar{\omega}_2$ since the $\bar{\omega}_1$ transition corresponds to a ${\omega}_i$ with the smaller imaginary part (the location of two peaks in the transmission are at $\Re\bar{\omega}_1=0.893$ and $\Re\bar{\omega}_2=1.411$ respectively which show small deviations from the real parts of the poles in the open system, $\Re\omega_i=0.898$ and $\Re\omega_i=1.384$.) Thus, Fig. (\ref{Transmissionfgk8}) shows that for a choice of physical parameters which leads to the imaginary parts of the eigenvalues of the effective Hamiltonian being small, we observe sharp transitions in the scattering spectrum in the infinite medium and these transitions occur close to the real parts of the eigenvalues of the effective Hamiltonian. 

A choice of $k=2$ N/m and $k_c=0.5$ N/m leads to $\omega_i$ with relatively large imaginary parts. Relevant $\omega_i$ and $\omega_i^N$ for this set of parameters are summarized in Table (\ref{table:Eigopenk15}) and the corresponding scattering spectrum is shown in Fig. (\ref{Transmissionfgk8}c). In the problem, we note that there are no sharp transitions in the scattering spectrum and this is a direct consequence of the relatively large magnitude of the imaginary parts of $\omega_i$. The first peak/dip in this problem occurs at $\bar{\omega}_1\approx 0.9$ which is well separated in the frequency domain from both $\Re \omega_i=0.734$ and $\omega_i^N=0.626$. The second scattering peak/dip is similarly mild and occurs at $\bar{\omega}_2\approx 1.5$.

\subsection{Case of $\bar{\omega}_i\approx\bar{\omega}_i^N$}

\begin{figure}[htp]
\centering
\includegraphics[scale=.5]{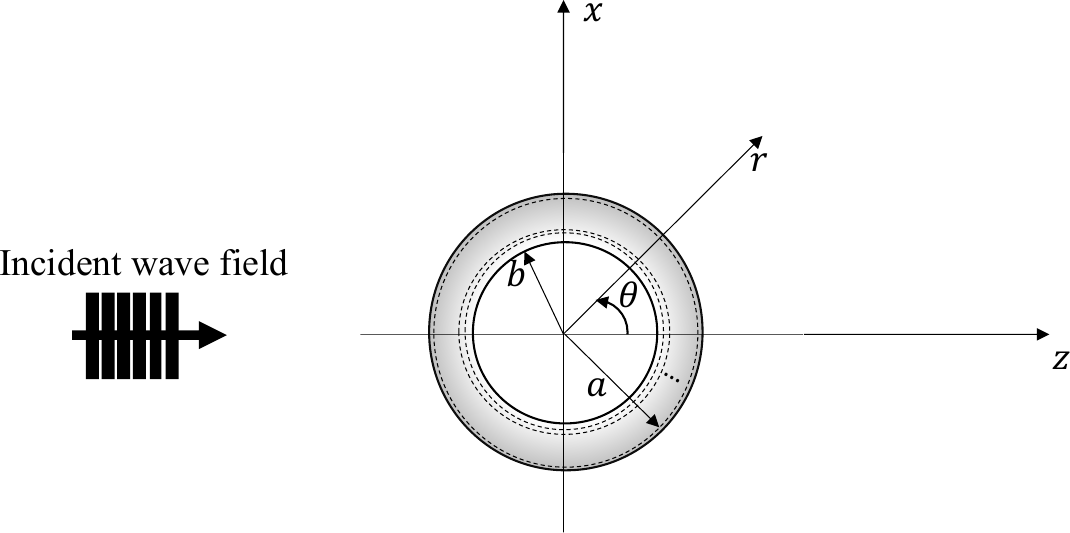}
\caption{An elastic sphere immersed in fluid illuminated by an in-plane acoustic wave field.}
\label{sphere}
\end{figure}

An interesting consequence of Eq. (\ref{Neumann3}) is that if there are eigenvalues of $\boldsymbol{H}_N$ in the low frequency regime then one can expect scattering resonances close to them. We show this by considering a 3D problem of a spherical elastic object illuminated by acoustic waves (Fig. \ref{sphere}). 
We provide the basic equations of the problem here with more details in classical references~\cite{scandrett2002scattering,hu1954general,ding1996natural}.
The time-independent form of the wave equation in the surrounding medium is:
\begin{equation}
    \label{wave}
    \left(\boldsymbol{\nabla}^2+q^2\right)\varphi(\boldsymbol{r})=0
\end{equation}
where $q=\omega/c$ and $\varphi(\boldsymbol{r})$ is the velocity potential field. The time-independent velocity can be expressed as $\boldsymbol{v}(\boldsymbol{r})=-\nabla \varphi(\boldsymbol{r})$.  
Expanding the incident and causal scattered fields in spherical harmonics:
\begin{align}
\label{phiincscc}
    \nonumber \varphi_{inc}(r,\theta,\omega,t)=\varphi_0\sum_{n=0}^{\infty}\left(2n+1\right)i^n j_n(qr)P_n(\cos \theta) e^{-i\omega t}\\
    \varphi_{sc}(r,\theta,\omega,t)=\varphi_0\sum_{n=0}^{\infty}A_n\left(2n+1\right)i^n h_n^{(1)}(qr)P_n(\cos \theta) e^{-i\omega t}
\end{align}
In the above, $\varphi_{0}$ is the incident velocity potential amplitude, $A_n$ are modal scattering coefficients, $P_n(\cos \theta)$ are the Legendre polynomials, $j_n(qr)$ is the spherical Bessel function and $h_n^{(1)}(qr)$ is the spherical Hankel function~\cite{mathhandbook}. For the spherically isotropic elastic shell, the governing equations of motion, constitutive equations and the strain-displacement relations are \cite{scandrett2002scattering}:

\begin{equation}
    \label{EOMsphere}
    \begin{aligned} \boldsymbol{\Gamma}\boldsymbol{\sigma}=\rho\ddot{\boldsymbol{u}},\quad \boldsymbol{\sigma}=\boldsymbol{CS}, \quad
    \boldsymbol{S}=\boldsymbol{H}\boldsymbol{u}
    \end{aligned}
\end{equation}
where, $\boldsymbol{u}$ is the displacement vector, $\boldsymbol{\sigma}$ is the stress vector, $\boldsymbol{S}$ is the strain vector and $\boldsymbol{C}$ is the stiffness matrix. The matrices $\boldsymbol{\Gamma}$ and $\boldsymbol{H}$ are provided in classical references~\cite{hu1954general,ding1996natural}. 

The above problem can be solved as either a scattering problem or a vibration problem. For the former, we can apply a continuity of pressure and normal velocity at $r=a$, whereas for vibration problem we can apply either free or fixed boundary conditions at $r=a$.

\begin{figure}[htp]
\centering
\includegraphics[scale=.122]{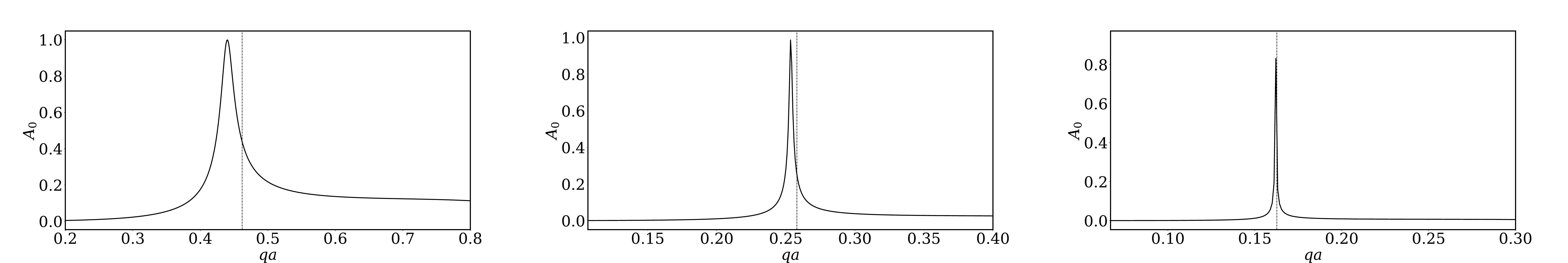}
\caption{Mode zero scattering coefficient vs. $qa$ for an elastic spherical shell with $E=20.7$ GPa, and (a) $\rho=1962$ kg/$m^3$, (b) $\rho=6280$ kg/$m^3$, (c) $\rho=15700$ kg/$m^3$ immersed in a fluid with $c_f=1800$ m/s and $\rho_f=8\times10^3$ kg/$m^3$. }
\label{scc0}
\end{figure}

Based upon earlier discussion, in this problem, we expect to see that in the small frequency regime, the scattering resonances should happen close to the eigenfrequencies corresponding to the scatterer under free boundary condition. The low frequency regime for the scatterer is characterized by or $e^{iqa}\approx 1$. Taking the low frequency limit usually considered as $qa\approx 0.1$, we expect that when the vibrational frequency of the scatterer (under free boundary condition) falls in this regime, then we should expect a close correspondence between it and the scattering resonance location. Fig. (\ref{scc0}) shows the scattering coefficient $A_0$, for a spherical shell with three different densities. In these figures, the vertical line gives the location of the first eigenfrequency of the scatterer under free boundary condition. For $\rho=15700$ kg/$m^3$, Fig. (\ref{scc0}. c) shows the condition under which the first vibrational frequency occurs in the low frequency regime and we note that it is very close to the location of the scattering resonance. Sub-figures (a) and (b) are plotted for lower densities which corresponds to peaks further away from the location of first natural frequency. Thus, it is clear that the general lessons learned from the analysis of Eq. (\ref{Neumann3}) may be applicable to more complicated scattering systems in higher dimensions as well.

\section{Conclusions}
In this paper, we have adapted a projection operator formalism from quantum mechanics for the study of mechanical wave scattering problems. The formalism allows us to derive insight into the physics of a scattering problem by looking at the problem as an interaction of a central system (finite dimensional subspace) with its environment (infinite dimensional subspace), instead of treating the problem as a single infinite dimensional problem (scatterer+environment). We showed that the projection operator formalism allows us to calculate the effective Hamiltonian of the open system, and that it is composed of a Hermitian operator corresponding to the finite scatterer isolated from its environment (zero Dirichlet boundary conditions), and a non-Hermitian contribution in the form of another operator representing the effect of its interaction with the environment. We also showed that the eigenvalues of the effective Hamiltonian closely predict the locations of the peaks/dips in the scattering spectrum of the infinite problem. In numerical results and discussion, we brought in findings from the preceding sections and showed that the coupling to the environment along with the stiffness in the scatterer can change the imaginary parts of the eigenvalues of the effective Hamiltonian. This directly affects the sharpness of the transitions in the scattering spectrum and we explicitly showed this phenomenon by solving the scattering problem for two specific cases.

The introduced projection operator formalism is general and applicable to both higher dimensional problems and to multi channel problems (multiple scattered modes in the environment). Teasing out these connections will likely lead to the development of straightforward numerical algorithms for implementing the projection operator formalism to arbitrarily complex problems. The viewpoint and machinery of open systems can also be used to better design scatterers, as it neatly separates out what is inherent to the scatterer ($H_0$) from what emerges due to the interplay between the scatterer and its environment ($\Sigma$). These ideas will be considered in future works.

\section*{Acknowledgments}
AS acknowledges NSF support for this work through grant \#1825969 to the Illinois Institute of Technology, Chicago.

\section*{Appendix}

\subsection{Details on the quadratic eigenvalue problem used in Section \ref{NR}}

In this appendix we describe the matrices and equations used in section \ref{NR}. The equations of motion of all the masses in Fig. (\ref{system1}) can be described as a system of infinite number of time independent coupled equations as:
\begin{equation}
\label{Ex1}
\left\{
\begin{array}{lcl}
  -\omega^2u_0=k_c(u_{1}-u_0)+k(u_{-1}-u_0)+k_{d_0}(u_{d_0}-u_0)  \\
  -\omega^2u_1=k_c(u_{0}-u_1)+k(u_{2}-u_1)+k_{d_1}(u_{d_1}-u_1)\\
  -\omega^2u_{d_0}=k_{d_0}(u_0-u_{d_0})\\
  -\omega^2u_{d_1}=k_{d_1}(u_1-u_{d_1})\\
  -\omega^2u_{n}=k(u_{n-1}+u_{n+1}-2u_{n}) \quad ; \quad n\neq0,1,d_0,d_1\\

\end{array}\right.
\end{equation}

Considering the relationships $u_{-1}=u_0e^{iqL}$ and $u_{2}=u_1e^{iqL}$ and also the dispersion relationship $\omega^2=2k(1-\cos{(qL)})$ for the two semi infinite branches connected to sites $0$ and $2$ we can write the above system of equations as:
\begin{equation}
\label{EOM2res}
    \begin{pmatrix}
    -k_c+k(e^{iqL}-1)-k_{d_0} & k_c & k_{d_0} & 0\\
    k_c& -k_c+k(e^{iqL}-1)-k_{d_1} & 0 & k_{d_1}\\
    k_{d_0} & 0 & -k_{d_0} & 0\\
    0 & k_{d_1} & 0 & -k_{d_1}
    \end{pmatrix}\boldsymbol{u}=
    -2k(1-\cos{(qL)})\boldsymbol{u}
\end{equation}
where $\boldsymbol{u}=\left(u_0,u_1,u_{d_0},u_{d_1}\right)^T$. By considering $\beta=e^{iqL}$, Eq.(\ref{EOM2res}) can be written as:
\begin{equation}
    \left(\boldsymbol{H}_0+\beta \boldsymbol{H}_1\right)\boldsymbol{u}=-2k\left(1-\frac{1}{2}\left(\beta+\frac{1}{\beta}\right)\right)\boldsymbol{u}
\end{equation}

Now, we convert this relationship to a quadratic eigenvalue problem, $\beta^2M+\beta C+K=0$, where matrices $M$, $C$, and $K$ are:
\begin{eqnarray}
M=\begin{pmatrix}
0& 0 & 0 & 0 \\
0& 0 & 0 & 0 \\
0& 0& -k& 0& \\
0& 0& 0& -k& 
\end{pmatrix}\\
C= \begin{pmatrix}
    -k_c+k-k_{d_0} & k_c & k_{d_0} & 0 \\
    k_c& -k_c+k-k_{d_1} & 0 & k_{d_1} \\
    k_{d_0} & 0 & 2k-k_{d_0} & 0 \\
    0 & k_{d_1} & 0 & -k_{d_1}+2k 
\end{pmatrix}\\
K=\begin{pmatrix}
-k& 0 & 0 & 0 \\
0& -k& 0 & 0 \\
0& 0 & -k & 0 \\
0& 0 & 0 & -k 
\end{pmatrix}
\end{eqnarray}

\section*{References}


\begin{thebibliography}{90}%
\makeatletter
\providecommand \@ifxundefined [1]{%
 \@ifx{#1\undefined}
}%
\providecommand \@ifnum [1]{%
 \ifnum #1\expandafter \@firstoftwo
 \else \expandafter \@secondoftwo
 \fi
}%
\providecommand \@ifx [1]{%
 \ifx #1\expandafter \@firstoftwo
 \else \expandafter \@secondoftwo
 \fi
}%
\providecommand \natexlab [1]{#1}%
\providecommand \enquote  [1]{``#1''}%
\providecommand \bibnamefont  [1]{#1}%
\providecommand \bibfnamefont [1]{#1}%
\providecommand \citenamefont [1]{#1}%
\providecommand \href@noop [0]{\@secondoftwo}%
\providecommand \href [0]{\begingroup \@sanitize@url \@href}%
\providecommand \@href[1]{\@@startlink{#1}\@@href}%
\providecommand \@@href[1]{\endgroup#1\@@endlink}%
\providecommand \@sanitize@url [0]{\catcode `\\12\catcode `\$12\catcode
  `\&12\catcode `\#12\catcode `\^12\catcode `\_12\catcode `\%12\relax}%
\providecommand \@@startlink[1]{}%
\providecommand \@@endlink[0]{}%
\providecommand \url  [0]{\begingroup\@sanitize@url \@url }%
\providecommand \@url [1]{\endgroup\@href {#1}{\urlprefix }}%
\providecommand \urlprefix  [0]{URL }%
\providecommand \Eprint [0]{\href }%
\providecommand \doibase [0]{http://dx.doi.org/}%
\providecommand \selectlanguage [0]{\@gobble}%
\providecommand \bibinfo  [0]{\@secondoftwo}%
\providecommand \bibfield  [0]{\@secondoftwo}%
\providecommand \translation [1]{[#1]}%
\providecommand \BibitemOpen [0]{}%
\providecommand \bibitemStop [0]{}%
\providecommand \bibitemNoStop [0]{.\EOS\space}%
\providecommand \EOS [0]{\spacefactor3000\relax}%
\providecommand \BibitemShut  [1]{\csname bibitem#1\endcsname}%
\let\auto@bib@innerbib\@empty
\bibitem [{\citenamefont {Rayleigh}(1917)}]{rayleigh1917reflection}%
  \BibitemOpen
  \bibfield  {author} {\bibinfo {author} {\bibfnamefont {Lord}\ \bibnamefont
  {Rayleigh}},\ }\bibfield  {title} {\enquote {\bibinfo {title} {{On the
  reflection of light from a regularly stratified medium}},}\ }\href@noop {}
  {\bibfield  {journal} {\bibinfo  {journal} {Proceedings of the Royal Society
  of London. Series A, Containing Papers of a Mathematical and Physical
  Character}\ }\textbf {\bibinfo {volume} {93}},\ \bibinfo {pages} {565--577}
  (\bibinfo {year} {1917})}\BibitemShut {NoStop}%
\bibitem [{\citenamefont {King}(1934)}]{King}%
  \BibitemOpen
  \bibfield  {author} {\bibinfo {author} {\bibfnamefont {Louis~Vessot}\
  \bibnamefont {King}},\ }\bibfield  {title} {\enquote {\bibinfo {title} {On
  the acoustic radiation pressure on spheres},}\ }\href {\doibase
  10.1098/rspa.1934.0215} {\bibfield  {journal} {\bibinfo  {journal}
  {Proceedings of the Royal Society of London. Series A - Mathematical and
  Physical Sciences}\ }\textbf {\bibinfo {volume} {147}},\ \bibinfo {pages}
  {212--240} (\bibinfo {year} {1934})},\ \Eprint
  {http://arxiv.org/abs/https://royalsocietypublishing.org/doi/pdf/10.1098/rspa.1934.0215}
  {https://royalsocietypublishing.org/doi/pdf/10.1098/rspa.1934.0215}
  \BibitemShut {NoStop}%
\bibitem [{\citenamefont {Willis}(1980)}]{willis1980polarization}%
  \BibitemOpen
  \bibfield  {author} {\bibinfo {author} {\bibfnamefont {JR}~\bibnamefont
  {Willis}},\ }\bibfield  {title} {\enquote {\bibinfo {title} {Polarization
  approach to the scattering of elastic waves—i. scattering by a single
  inclusion},}\ }\href@noop {} {\bibfield  {journal} {\bibinfo  {journal}
  {Journal of the Mechanics and Physics of Solids}\ }\textbf {\bibinfo {volume}
  {28}},\ \bibinfo {pages} {287--305} (\bibinfo {year} {1980})}\BibitemShut
  {NoStop}%
\bibitem [{\citenamefont {Eshelby}(1957)}]{eshelby1957determination}%
  \BibitemOpen
  \bibfield  {author} {\bibinfo {author} {\bibfnamefont {J~D}\ \bibnamefont
  {Eshelby}},\ }\bibfield  {title} {\enquote {\bibinfo {title} {{The
  determination of the elastic field of an ellipsoidal inclusion, and related
  problems}},}\ }\href@noop {} {\bibfield  {journal} {\bibinfo  {journal}
  {Proceedings of the Royal Society of London. Series A. Mathematical and
  Physical Sciences}\ }\textbf {\bibinfo {volume} {241}},\ \bibinfo {pages}
  {376} (\bibinfo {year} {1957})}\BibitemShut {NoStop}%
\bibitem [{\citenamefont {Westervelt}(1957)}]{westervelt1957acoustic}%
  \BibitemOpen
  \bibfield  {author} {\bibinfo {author} {\bibfnamefont {Peter~J}\ \bibnamefont
  {Westervelt}},\ }\bibfield  {title} {\enquote {\bibinfo {title} {Acoustic
  radiation pressure},}\ }\href@noop {} {\bibfield  {journal} {\bibinfo
  {journal} {The Journal of the Acoustical Society of America}\ }\textbf
  {\bibinfo {volume} {29}},\ \bibinfo {pages} {26--29} (\bibinfo {year}
  {1957})}\BibitemShut {NoStop}%
\bibitem [{\citenamefont {Faran~Jr}(1951)}]{faran1951sound}%
  \BibitemOpen
  \bibfield  {author} {\bibinfo {author} {\bibfnamefont {James~J}\ \bibnamefont
  {Faran~Jr}},\ }\bibfield  {title} {\enquote {\bibinfo {title} {Sound
  scattering by solid cylinders and spheres},}\ }\href@noop {} {\bibfield
  {journal} {\bibinfo  {journal} {The Journal of the acoustical society of
  America}\ }\textbf {\bibinfo {volume} {23}},\ \bibinfo {pages} {405--418}
  (\bibinfo {year} {1951})}\BibitemShut {NoStop}%
\bibitem [{\citenamefont {Krein}(1956)}]{krein1956theory}%
  \BibitemOpen
  \bibfield  {author} {\bibinfo {author} {\bibfnamefont {Mark~G}\ \bibnamefont
  {Krein}},\ }\bibfield  {title} {\enquote {\bibinfo {title} {The theory of
  accelerants and s-matrices of canonical differential systems},}\ }\href@noop
  {} {\bibfield  {journal} {\bibinfo  {journal} {Doklady Akademii Nauk SSSR}\
  }\textbf {\bibinfo {volume} {111}},\ \bibinfo {pages} {1167--1170} (\bibinfo
  {year} {1956})}\BibitemShut {NoStop}%
\bibitem [{\citenamefont {Krein}\ and\ \citenamefont
  {Birman}(1962)}]{krein1962theory}%
  \BibitemOpen
  \bibfield  {author} {\bibinfo {author} {\bibfnamefont {Mark~Grigorevich}\
  \bibnamefont {Krein}}\ and\ \bibinfo {author} {\bibfnamefont
  {MS}~\bibnamefont {Birman}},\ }\bibfield  {title} {\enquote {\bibinfo {title}
  {On the theory of wave operators and scattering operators},}\ }in\ \href@noop
  {} {\emph {\bibinfo {booktitle} {Dokl. Akad. Nauk SSSR}}},\ Vol.\ \bibinfo
  {volume} {144}\ (\bibinfo {year} {1962})\ pp.\ \bibinfo {pages}
  {740--744}\BibitemShut {NoStop}%
\bibitem [{\citenamefont {Milton}\ \emph {et~al.}(2006)\citenamefont {Milton},
  \citenamefont {Briane},\ and\ \citenamefont {Willis}}]{milton2006cloaking}%
  \BibitemOpen
  \bibfield  {author} {\bibinfo {author} {\bibfnamefont {G~W}\ \bibnamefont
  {Milton}}, \bibinfo {author} {\bibfnamefont {M}~\bibnamefont {Briane}}, \
  and\ \bibinfo {author} {\bibfnamefont {J~R}\ \bibnamefont {Willis}},\
  }\bibfield  {title} {\enquote {\bibinfo {title} {{On cloaking for elasticity
  and physical equations with a transformation invariant form}},}\ }\href@noop
  {} {\bibfield  {journal} {\bibinfo  {journal} {New journal of physics}\
  }\textbf {\bibinfo {volume} {8}},\ \bibinfo {pages} {248} (\bibinfo {year}
  {2006})}\BibitemShut {NoStop}%
\bibitem [{\citenamefont {Wolf}\ and\ \citenamefont
  {Habashy}(1993)}]{wolf1993invisible}%
  \BibitemOpen
  \bibfield  {author} {\bibinfo {author} {\bibfnamefont {Emil}\ \bibnamefont
  {Wolf}}\ and\ \bibinfo {author} {\bibfnamefont {Tarek}\ \bibnamefont
  {Habashy}},\ }\bibfield  {title} {\enquote {\bibinfo {title} {Invisible
  bodies and uniqueness of the inverse scattering problem},}\ }\href@noop {}
  {\bibfield  {journal} {\bibinfo  {journal} {Journal of Modern Optics}\
  }\textbf {\bibinfo {volume} {40}},\ \bibinfo {pages} {785--792} (\bibinfo
  {year} {1993})}\BibitemShut {NoStop}%
\bibitem [{\citenamefont {Chen}\ and\ \citenamefont
  {Chan}(2007)}]{chen2007acoustic}%
  \BibitemOpen
  \bibfield  {author} {\bibinfo {author} {\bibfnamefont {H}~\bibnamefont
  {Chen}}\ and\ \bibinfo {author} {\bibfnamefont {C~T}\ \bibnamefont {Chan}},\
  }\bibfield  {title} {\enquote {\bibinfo {title} {{Acoustic cloaking in three
  dimensions using acoustic metamaterials}},}\ }\href@noop {} {\bibfield
  {journal} {\bibinfo  {journal} {Applied physics letters}\ }\textbf {\bibinfo
  {volume} {91}},\ \bibinfo {pages} {183518} (\bibinfo {year}
  {2007})}\BibitemShut {NoStop}%
\bibitem [{\citenamefont {Cummer}\ \emph {et~al.}(2008)\citenamefont {Cummer},
  \citenamefont {Popa}, \citenamefont {Schurig}, \citenamefont {Smith},
  \citenamefont {Pendry}, \citenamefont {Rahm},\ and\ \citenamefont
  {Starr}}]{cummer2008scattering}%
  \BibitemOpen
  \bibfield  {author} {\bibinfo {author} {\bibfnamefont {Steven~A}\
  \bibnamefont {Cummer}}, \bibinfo {author} {\bibfnamefont {Bogdan-Ioan}\
  \bibnamefont {Popa}}, \bibinfo {author} {\bibfnamefont {David}\ \bibnamefont
  {Schurig}}, \bibinfo {author} {\bibfnamefont {David~R}\ \bibnamefont
  {Smith}}, \bibinfo {author} {\bibfnamefont {John}\ \bibnamefont {Pendry}},
  \bibinfo {author} {\bibfnamefont {Marco}\ \bibnamefont {Rahm}}, \ and\
  \bibinfo {author} {\bibfnamefont {Anthony}\ \bibnamefont {Starr}},\
  }\bibfield  {title} {\enquote {\bibinfo {title} {Scattering theory derivation
  of a 3d acoustic cloaking shell},}\ }\href@noop {} {\bibfield  {journal}
  {\bibinfo  {journal} {Physical review letters}\ }\textbf {\bibinfo {volume}
  {100}},\ \bibinfo {pages} {024301} (\bibinfo {year} {2008})}\BibitemShut
  {NoStop}%
\bibitem [{\citenamefont {Norris}(2008)}]{norris2008acoustic}%
  \BibitemOpen
  \bibfield  {author} {\bibinfo {author} {\bibfnamefont {A~N}\ \bibnamefont
  {Norris}},\ }\bibfield  {title} {\enquote {\bibinfo {title} {{Acoustic
  cloaking theory}},}\ }\href@noop {} {\bibfield  {journal} {\bibinfo
  {journal} {Proceedings of the Royal Society A: Mathematical, Physical and
  Engineering Science}\ }\textbf {\bibinfo {volume} {464}},\ \bibinfo {pages}
  {2411} (\bibinfo {year} {2008})}\BibitemShut {NoStop}%
\bibitem [{\citenamefont {Krasnok}\ and\ \citenamefont
  {Al{\'{u}}}(2018)}]{Krasnok_2018}%
  \BibitemOpen
  \bibfield  {author} {\bibinfo {author} {\bibfnamefont {Alex}\ \bibnamefont
  {Krasnok}}\ and\ \bibinfo {author} {\bibfnamefont {Andrea}\ \bibnamefont
  {Al{\'{u}}}},\ }\bibfield  {title} {\enquote {\bibinfo {title} {Embedded
  scattering eigenstates using resonant metasurfaces},}\ }\href {\doibase
  10.1088/2040-8986/aac1d6} {\bibfield  {journal} {\bibinfo  {journal} {Journal
  of Optics}\ }\textbf {\bibinfo {volume} {20}},\ \bibinfo {pages} {064002}
  (\bibinfo {year} {2018})}\BibitemShut {NoStop}%
\bibitem [{\citenamefont {Assouar}\ \emph {et~al.}(2018)\citenamefont
  {Assouar}, \citenamefont {Liang}, \citenamefont {Wu}, \citenamefont {Li},
  \citenamefont {Cheng},\ and\ \citenamefont {Jing}}]{assouar2018acoustic}%
  \BibitemOpen
  \bibfield  {author} {\bibinfo {author} {\bibfnamefont {Badreddine}\
  \bibnamefont {Assouar}}, \bibinfo {author} {\bibfnamefont {Bin}\ \bibnamefont
  {Liang}}, \bibinfo {author} {\bibfnamefont {Ying}\ \bibnamefont {Wu}},
  \bibinfo {author} {\bibfnamefont {Yong}\ \bibnamefont {Li}}, \bibinfo
  {author} {\bibfnamefont {Jian-Chun}\ \bibnamefont {Cheng}}, \ and\ \bibinfo
  {author} {\bibfnamefont {Yun}\ \bibnamefont {Jing}},\ }\bibfield  {title}
  {\enquote {\bibinfo {title} {Acoustic metasurfaces},}\ }\href@noop {}
  {\bibfield  {journal} {\bibinfo  {journal} {Nature Reviews Materials}\
  }\textbf {\bibinfo {volume} {3}},\ \bibinfo {pages} {460--472} (\bibinfo
  {year} {2018})}\BibitemShut {NoStop}%
\bibitem [{\citenamefont {Chen}\ \emph {et~al.}(2016)\citenamefont {Chen},
  \citenamefont {Taylor},\ and\ \citenamefont {Yu}}]{chen2016review}%
  \BibitemOpen
  \bibfield  {author} {\bibinfo {author} {\bibfnamefont {Hou-Tong}\
  \bibnamefont {Chen}}, \bibinfo {author} {\bibfnamefont {Antoinette~J}\
  \bibnamefont {Taylor}}, \ and\ \bibinfo {author} {\bibfnamefont {Nanfang}\
  \bibnamefont {Yu}},\ }\bibfield  {title} {\enquote {\bibinfo {title} {A
  review of metasurfaces: physics and applications},}\ }\href@noop {}
  {\bibfield  {journal} {\bibinfo  {journal} {Reports on progress in physics}\
  }\textbf {\bibinfo {volume} {79}},\ \bibinfo {pages} {076401} (\bibinfo
  {year} {2016})}\BibitemShut {NoStop}%
\bibitem [{\citenamefont {Cahill}\ \emph {et~al.}(2014)\citenamefont {Cahill},
  \citenamefont {Braun}, \citenamefont {Chen}, \citenamefont {Clarke},
  \citenamefont {Fan}, \citenamefont {Goodson}, \citenamefont {Keblinski},
  \citenamefont {King}, \citenamefont {Mahan}, \citenamefont {Majumdar},
  \citenamefont {Maris}, \citenamefont {Phillpot}, \citenamefont {Pop},\ and\
  \citenamefont {Shi}}]{Nanoscalethermalreview}%
  \BibitemOpen
  \bibfield  {author} {\bibinfo {author} {\bibfnamefont {David~G.}\
  \bibnamefont {Cahill}}, \bibinfo {author} {\bibfnamefont {Paul~V.}\
  \bibnamefont {Braun}}, \bibinfo {author} {\bibfnamefont {Gang}\ \bibnamefont
  {Chen}}, \bibinfo {author} {\bibfnamefont {David~R.}\ \bibnamefont {Clarke}},
  \bibinfo {author} {\bibfnamefont {Shanhui}\ \bibnamefont {Fan}}, \bibinfo
  {author} {\bibfnamefont {Kenneth~E.}\ \bibnamefont {Goodson}}, \bibinfo
  {author} {\bibfnamefont {Pawel}\ \bibnamefont {Keblinski}}, \bibinfo {author}
  {\bibfnamefont {William~P.}\ \bibnamefont {King}}, \bibinfo {author}
  {\bibfnamefont {Gerald~D.}\ \bibnamefont {Mahan}}, \bibinfo {author}
  {\bibfnamefont {Arun}\ \bibnamefont {Majumdar}}, \bibinfo {author}
  {\bibfnamefont {Humphrey~J.}\ \bibnamefont {Maris}}, \bibinfo {author}
  {\bibfnamefont {Simon~R.}\ \bibnamefont {Phillpot}}, \bibinfo {author}
  {\bibfnamefont {Eric}\ \bibnamefont {Pop}}, \ and\ \bibinfo {author}
  {\bibfnamefont {Li}~\bibnamefont {Shi}},\ }\bibfield  {title} {\enquote
  {\bibinfo {title} {Nanoscale thermal transport. ii. 2003–2012},}\ }\href
  {\doibase 10.1063/1.4832615} {\bibfield  {journal} {\bibinfo  {journal}
  {Applied Physics Reviews}\ }\textbf {\bibinfo {volume} {1}},\ \bibinfo
  {pages} {011305} (\bibinfo {year} {2014})},\ \Eprint
  {http://arxiv.org/abs/https://doi.org/10.1063/1.4832615}
  {https://doi.org/10.1063/1.4832615} \BibitemShut {NoStop}%
\bibitem [{\citenamefont {Zhang}\ \emph {et~al.}(2007)\citenamefont {Zhang},
  \citenamefont {Fisher},\ and\ \citenamefont {Mingo}}]{AGFNanoscalePhonon}%
  \BibitemOpen
  \bibfield  {author} {\bibinfo {author} {\bibfnamefont {W.}~\bibnamefont
  {Zhang}}, \bibinfo {author} {\bibfnamefont {T.~S.}\ \bibnamefont {Fisher}}, \
  and\ \bibinfo {author} {\bibfnamefont {N.}~\bibnamefont {Mingo}},\ }\bibfield
   {title} {\enquote {\bibinfo {title} {The atomistic green's function method:
  An efficient simulation approach for nanoscale phonon transport},}\ }\href
  {\doibase 10.1080/10407790601144755} {\bibfield  {journal} {\bibinfo
  {journal} {Numerical Heat Transfer, Part B: Fundamentals}\ }\textbf {\bibinfo
  {volume} {51}},\ \bibinfo {pages} {333--349} (\bibinfo {year} {2007})},\
  \Eprint {http://arxiv.org/abs/https://doi.org/10.1080/10407790601144755}
  {https://doi.org/10.1080/10407790601144755} \BibitemShut {NoStop}%
\bibitem [{\citenamefont {Ong}\ and\ \citenamefont
  {Zhang}(2015)}]{PhysRevB.91.174302}%
  \BibitemOpen
  \bibfield  {author} {\bibinfo {author} {\bibfnamefont {Zhun-Yong}\
  \bibnamefont {Ong}}\ and\ \bibinfo {author} {\bibfnamefont {Gang}\
  \bibnamefont {Zhang}},\ }\bibfield  {title} {\enquote {\bibinfo {title}
  {Efficient approach for modeling phonon transmission probability in nanoscale
  interfacial thermal transport},}\ }\href {\doibase
  10.1103/PhysRevB.91.174302} {\bibfield  {journal} {\bibinfo  {journal} {Phys.
  Rev. B}\ }\textbf {\bibinfo {volume} {91}},\ \bibinfo {pages} {174302}
  (\bibinfo {year} {2015})}\BibitemShut {NoStop}%
\bibitem [{\citenamefont {Hussein}\ \emph {et~al.}(2014)\citenamefont
  {Hussein}, \citenamefont {Leamy},\ and\ \citenamefont
  {Ruzzene}}]{MHusseinReview}%
  \BibitemOpen
  \bibfield  {author} {\bibinfo {author} {\bibfnamefont {Mahmoud~I.}\
  \bibnamefont {Hussein}}, \bibinfo {author} {\bibfnamefont {Michael~J.}\
  \bibnamefont {Leamy}}, \ and\ \bibinfo {author} {\bibfnamefont {Massimo}\
  \bibnamefont {Ruzzene}},\ }\bibfield  {title} {\enquote {\bibinfo {title}
  {{Dynamics of Phononic Materials and Structures: Historical Origins, Recent
  Progress, and Future Outlook}},}\ }\href {\doibase 10.1115/1.4026911}
  {\bibfield  {journal} {\bibinfo  {journal} {Applied Mechanics Reviews}\
  }\textbf {\bibinfo {volume} {66}} (\bibinfo {year} {2014}),\
  10.1115/1.4026911},\ \bibinfo {note} {040802},\ \Eprint
  {http://arxiv.org/abs/https://asmedigitalcollection.asme.org/appliedmechanicsreviews/article-pdf/66/4/040802/6782437/amr\_066\_04\_040802.pdf}
  {https://asmedigitalcollection.asme.org/appliedmechanicsreviews/article-pdf/66/4/040802/6782437/amr\_066\_04\_040802.pdf}
  \BibitemShut {NoStop}%
\bibitem [{\citenamefont {Mokhtari}\ \emph {et~al.}(2020)\citenamefont
  {Mokhtari}, \citenamefont {Lu}, \citenamefont {Zhou}, \citenamefont
  {Amirkhizi},\ and\ \citenamefont {Srivastava}}]{mokhtari2020scattering}%
  \BibitemOpen
  \bibfield  {author} {\bibinfo {author} {\bibfnamefont {Amir~Ashkan}\
  \bibnamefont {Mokhtari}}, \bibinfo {author} {\bibfnamefont {Yan}\
  \bibnamefont {Lu}}, \bibinfo {author} {\bibfnamefont {Qiyuan}\ \bibnamefont
  {Zhou}}, \bibinfo {author} {\bibfnamefont {Alireza~V}\ \bibnamefont
  {Amirkhizi}}, \ and\ \bibinfo {author} {\bibfnamefont {Ankit}\ \bibnamefont
  {Srivastava}},\ }\bibfield  {title} {\enquote {\bibinfo {title} {Scattering
  of in-plane elastic waves at metamaterial interfaces},}\ }\href@noop {}
  {\bibfield  {journal} {\bibinfo  {journal} {International Journal of
  Engineering Science}\ }\textbf {\bibinfo {volume} {150}},\ \bibinfo {pages}
  {103278} (\bibinfo {year} {2020})}\BibitemShut {NoStop}%
\bibitem [{\citenamefont {Mokhtari}\ and\ \citenamefont
  {Srivastava}(2019)}]{mokhtari2019properties}%
  \BibitemOpen
  \bibfield  {author} {\bibinfo {author} {\bibfnamefont {Amir~Ashkan}\
  \bibnamefont {Mokhtari}}\ and\ \bibinfo {author} {\bibfnamefont {Ankit}\
  \bibnamefont {Srivastava}},\ }\bibfield  {title} {\enquote {\bibinfo {title}
  {On the properties of phononic eigenvalue problems},}\ }\href@noop {}
  {\bibfield  {journal} {\bibinfo  {journal} {arXiv preprint arXiv:1902.07144}\
  } (\bibinfo {year} {2019})}\BibitemShut {NoStop}%
\bibitem [{\citenamefont {Baz}(2010)}]{baz2010active}%
  \BibitemOpen
  \bibfield  {author} {\bibinfo {author} {\bibfnamefont {Amr~M}\ \bibnamefont
  {Baz}},\ }\bibfield  {title} {\enquote {\bibinfo {title} {An active acoustic
  metamaterial with tunable effective density},}\ }\href@noop {} {\bibfield
  {journal} {\bibinfo  {journal} {Journal of vibration and acoustics}\ }\textbf
  {\bibinfo {volume} {132}} (\bibinfo {year} {2010})}\BibitemShut {NoStop}%
\bibitem [{\citenamefont {Skudrzyk}(2012)}]{skudrzyk2012foundations}%
  \BibitemOpen
  \bibfield  {author} {\bibinfo {author} {\bibfnamefont {Eugen}\ \bibnamefont
  {Skudrzyk}},\ }\href@noop {} {\emph {\bibinfo {title} {The foundations of
  acoustics: basic mathematics and basic acoustics}}}\ (\bibinfo  {publisher}
  {Springer Science \& Business Media},\ \bibinfo {year} {2012})\BibitemShut
  {NoStop}%
\bibitem [{\citenamefont {Roach}(2009)}]{roach2009wave}%
  \BibitemOpen
  \bibfield  {author} {\bibinfo {author} {\bibfnamefont {Gary~Francis}\
  \bibnamefont {Roach}},\ }\href@noop {} {\emph {\bibinfo {title} {Wave
  scattering by time-dependent perturbations}}}\ (\bibinfo  {publisher}
  {Princeton University Press},\ \bibinfo {year} {2009})\BibitemShut {NoStop}%
\bibitem [{\citenamefont {Ashida}\ \emph {et~al.}(2020)\citenamefont {Ashida},
  \citenamefont {Gong},\ and\ \citenamefont {Ueda}}]{NonHermitianPhys}%
  \BibitemOpen
  \bibfield  {author} {\bibinfo {author} {\bibfnamefont {Yuto}\ \bibnamefont
  {Ashida}}, \bibinfo {author} {\bibfnamefont {Zongping}\ \bibnamefont {Gong}},
  \ and\ \bibinfo {author} {\bibfnamefont {Masahito}\ \bibnamefont {Ueda}},\
  }\bibfield  {title} {\enquote {\bibinfo {title} {Non-hermitian physics},}\
  }\href {\doibase 10.1080/00018732.2021.1876991} {\bibfield  {journal}
  {\bibinfo  {journal} {Advances in Physics}\ }\textbf {\bibinfo {volume}
  {69}},\ \bibinfo {pages} {249--435} (\bibinfo {year} {2020})},\ \Eprint
  {http://arxiv.org/abs/https://doi.org/10.1080/00018732.2021.1876991}
  {https://doi.org/10.1080/00018732.2021.1876991} \BibitemShut {NoStop}%
\bibitem [{\citenamefont {Kamenetskii}\ \emph
  {et~al.}(2018{\natexlab{a}})\citenamefont {Kamenetskii}, \citenamefont
  {Sadreev},\ and\ \citenamefont {Miroshnichenko}}]{fanobook}%
  \BibitemOpen
  \bibfield  {author} {\bibinfo {author} {\bibfnamefont {Eugene}\ \bibnamefont
  {Kamenetskii}}, \bibinfo {author} {\bibfnamefont {Almas}\ \bibnamefont
  {Sadreev}}, \ and\ \bibinfo {author} {\bibfnamefont {Andrey}\ \bibnamefont
  {Miroshnichenko}},\ }\bibfield  {title} {\enquote {\bibinfo {title} {Fano
  resonances in optics and microwaves},}\ }\href@noop {} {\bibfield  {journal}
  {\bibinfo  {journal} {Physics and Applications Springer Series in Optical
  Sciences}\ }\textbf {\bibinfo {volume} {219}} (\bibinfo {year}
  {2018}{\natexlab{a}})}\BibitemShut {NoStop}%
\bibitem [{\citenamefont {Cohen-Tannoudji}\ \emph {et~al.}(2006)\citenamefont
  {Cohen-Tannoudji}, \citenamefont {Diu}, \citenamefont {Laloe},\ and\
  \citenamefont {Dui}}]{cohen2006quantum}%
  \BibitemOpen
  \bibfield  {author} {\bibinfo {author} {\bibfnamefont {Claude}\ \bibnamefont
  {Cohen-Tannoudji}}, \bibinfo {author} {\bibfnamefont {Bernard}\ \bibnamefont
  {Diu}}, \bibinfo {author} {\bibfnamefont {Frank}\ \bibnamefont {Laloe}}, \
  and\ \bibinfo {author} {\bibfnamefont {Bernard}\ \bibnamefont {Dui}},\
  }\href@noop {} {\enquote {\bibinfo {title} {Quantum mechanics (2 vol.
  set)},}\ } (\bibinfo {year} {2006})\BibitemShut {NoStop}%
\bibitem [{\citenamefont {Kamenetskii}\ \emph
  {et~al.}(2018{\natexlab{b}})\citenamefont {Kamenetskii}, \citenamefont
  {Sadreev},\ and\ \citenamefont {Miroshnichenko}}]{kamenetskii2018fano}%
  \BibitemOpen
  \bibfield  {author} {\bibinfo {author} {\bibfnamefont {Eugene}\ \bibnamefont
  {Kamenetskii}}, \bibinfo {author} {\bibfnamefont {Almas}\ \bibnamefont
  {Sadreev}}, \ and\ \bibinfo {author} {\bibfnamefont {Andrey}\ \bibnamefont
  {Miroshnichenko}},\ }\bibfield  {title} {\enquote {\bibinfo {title} {Fano
  resonances in optics and microwaves},}\ }\href@noop {} {\bibfield  {journal}
  {\bibinfo  {journal} {Physics and Applications Springer Series in Optical
  Sciences}\ }\textbf {\bibinfo {volume} {219}} (\bibinfo {year}
  {2018}{\natexlab{b}})}\BibitemShut {NoStop}%
\bibitem [{\citenamefont {Livsic}(2008)}]{Livsic}%
  \BibitemOpen
  \bibfield  {author} {\bibinfo {author} {\bibfnamefont {Moshe~S.}\
  \bibnamefont {Livsic}},\ }\bibfield  {title} {\enquote {\bibinfo {title}
  {Operators, oscillations, waves. open systems},}\ \ }(\bibinfo {year}
  {2008})\BibitemShut {NoStop}%
\bibitem [{\citenamefont {Garmon}\ \emph
  {et~al.}(2015{\natexlab{a}})\citenamefont {Garmon}, \citenamefont
  {Gianfreda},\ and\ \citenamefont {Hatano}}]{boundscattering}%
  \BibitemOpen
  \bibfield  {author} {\bibinfo {author} {\bibfnamefont {Savannah}\
  \bibnamefont {Garmon}}, \bibinfo {author} {\bibfnamefont {Mariagiovanna}\
  \bibnamefont {Gianfreda}}, \ and\ \bibinfo {author} {\bibfnamefont
  {Naomichi}\ \bibnamefont {Hatano}},\ }\bibfield  {title} {\enquote {\bibinfo
  {title} {Bound states, scattering states, and resonant states in
  $\mathcal{PT}$-symmetric open quantum systems},}\ }\href {\doibase
  10.1103/PhysRevA.92.022125} {\bibfield  {journal} {\bibinfo  {journal} {Phys.
  Rev. A}\ }\textbf {\bibinfo {volume} {92}},\ \bibinfo {pages} {022125}
  (\bibinfo {year} {2015}{\natexlab{a}})}\BibitemShut {NoStop}%
\bibitem [{\citenamefont {Moiseyev}(2011)}]{moiseyev2011non}%
  \BibitemOpen
  \bibfield  {author} {\bibinfo {author} {\bibfnamefont {Nimrod}\ \bibnamefont
  {Moiseyev}},\ }\href@noop {} {\emph {\bibinfo {title} {Non-Hermitian quantum
  mechanics}}}\ (\bibinfo  {publisher} {Cambridge University Press},\ \bibinfo
  {year} {2011})\BibitemShut {NoStop}%
\bibitem [{\citenamefont {Huang}\ \emph {et~al.}(2017)\citenamefont {Huang},
  \citenamefont {Shen}, \citenamefont {Min}, \citenamefont {Fan},\ and\
  \citenamefont {Veronis}}]{unidirectionalref}%
  \BibitemOpen
  \bibfield  {author} {\bibinfo {author} {\bibfnamefont {Yin}\ \bibnamefont
  {Huang}}, \bibinfo {author} {\bibfnamefont {Yuecheng}\ \bibnamefont {Shen}},
  \bibinfo {author} {\bibfnamefont {Changjun}\ \bibnamefont {Min}}, \bibinfo
  {author} {\bibfnamefont {Shanhui}\ \bibnamefont {Fan}}, \ and\ \bibinfo
  {author} {\bibfnamefont {Georgios}\ \bibnamefont {Veronis}},\ }\bibfield
  {title} {\enquote {\bibinfo {title} {Unidirectional reflectionless light
  propagation at exceptional points},}\ }\href {\doibase
  doi:10.1515/nanoph-2017-0019} {\bibfield  {journal} {\bibinfo  {journal}
  {Nanophotonics}\ }\textbf {\bibinfo {volume} {6}},\ \bibinfo {pages}
  {977--996} (\bibinfo {year} {2017})}\BibitemShut {NoStop}%
\bibitem [{\citenamefont {Kato}(1995)}]{LinearPerturbation}%
  \BibitemOpen
  \bibinfo {editor} {\bibfnamefont {Tosio}\ \bibnamefont {Kato}},\ ed.,\ \href
  {\doibase 10.1007/978-3-642-66282-9} {\emph {\bibinfo {title} {Perturbation
  Theory for Linear Operators}}}\ (\bibinfo  {publisher} {Springer-Verlag
  Berlin Heidelberg},\ \bibinfo {year} {1995})\BibitemShut {NoStop}%
\bibitem [{\citenamefont {Seyranian}(1993)}]{Sensitivityanalysis}%
  \BibitemOpen
  \bibfield  {author} {\bibinfo {author} {\bibfnamefont {Alexander~P.}\
  \bibnamefont {Seyranian}},\ }\bibfield  {title} {\enquote {\bibinfo {title}
  {Sensitivity analysis of multiple eigenvalues*},}\ }\href {\doibase
  10.1080/08905459308905189} {\bibfield  {journal} {\bibinfo  {journal}
  {Mechanics of Structures and Machines}\ }\textbf {\bibinfo {volume} {21}},\
  \bibinfo {pages} {261--284} (\bibinfo {year} {1993})},\ \Eprint
  {http://arxiv.org/abs/https://doi.org/10.1080/08905459308905189}
  {https://doi.org/10.1080/08905459308905189} \BibitemShut {NoStop}%
\bibitem [{\citenamefont {A.~P.~Seyranian}(1994)}]{multipleeigenvalue}%
  \BibitemOpen
  \bibfield  {author} {\bibinfo {author} {\bibfnamefont {N.~Olhoff}\
  \bibnamefont {A.~P.~Seyranian}, \bibfnamefont {E.~Lund}},\ }\bibfield
  {title} {\enquote {\bibinfo {title} {Multiple eigenvalues in structural
  optimization problems},}\ }\href {\doibase doi:10.1007/BF01742705} {\bibfield
   {journal} {\bibinfo  {journal} {Structural optimization}\ }\textbf {\bibinfo
  {volume} {8}},\ \bibinfo {pages} {207–227} (\bibinfo {year}
  {1994})}\BibitemShut {NoStop}%
\bibitem [{\citenamefont {Heiss}(2012)}]{heiss2012physics}%
  \BibitemOpen
  \bibfield  {author} {\bibinfo {author} {\bibfnamefont {WD}~\bibnamefont
  {Heiss}},\ }\bibfield  {title} {\enquote {\bibinfo {title} {The physics of
  exceptional points},}\ }\href@noop {} {\bibfield  {journal} {\bibinfo
  {journal} {Journal of Physics A: Mathematical and Theoretical}\ }\textbf
  {\bibinfo {volume} {45}},\ \bibinfo {pages} {444016} (\bibinfo {year}
  {2012})}\BibitemShut {NoStop}%
\bibitem [{\citenamefont {Heiss}\ and\ \citenamefont
  {Harney}(2001)}]{heiss2001chirality}%
  \BibitemOpen
  \bibfield  {author} {\bibinfo {author} {\bibfnamefont {WD}~\bibnamefont
  {Heiss}}\ and\ \bibinfo {author} {\bibfnamefont {HL}~\bibnamefont {Harney}},\
  }\bibfield  {title} {\enquote {\bibinfo {title} {The chirality of exceptional
  points},}\ }\href@noop {} {\bibfield  {journal} {\bibinfo  {journal} {The
  European Physical Journal D-Atomic, Molecular, Optical and Plasma Physics}\
  }\textbf {\bibinfo {volume} {17}},\ \bibinfo {pages} {149--151} (\bibinfo
  {year} {2001})}\BibitemShut {NoStop}%
\bibitem [{\citenamefont {Heiss}(2004)}]{heiss2004exceptional}%
  \BibitemOpen
  \bibfield  {author} {\bibinfo {author} {\bibfnamefont {WD}~\bibnamefont
  {Heiss}},\ }\bibfield  {title} {\enquote {\bibinfo {title} {Exceptional
  points of non-hermitian operators},}\ }\href@noop {} {\bibfield  {journal}
  {\bibinfo  {journal} {Journal of Physics A: Mathematical and General}\
  }\textbf {\bibinfo {volume} {37}},\ \bibinfo {pages} {2455} (\bibinfo {year}
  {2004})}\BibitemShut {NoStop}%
\bibitem [{\citenamefont {Lu}\ and\ \citenamefont
  {Srivastava}(2018)}]{LU2018100}%
  \BibitemOpen
  \bibfield  {author} {\bibinfo {author} {\bibfnamefont {Yan}\ \bibnamefont
  {Lu}}\ and\ \bibinfo {author} {\bibfnamefont {Ankit}\ \bibnamefont
  {Srivastava}},\ }\bibfield  {title} {\enquote {\bibinfo {title} {Level
  repulsion and band sorting in phononic crystals},}\ }\href {\doibase
  https://doi.org/10.1016/j.jmps.2017.10.021} {\bibfield  {journal} {\bibinfo
  {journal} {Journal of the Mechanics and Physics of Solids}\ }\textbf
  {\bibinfo {volume} {111}},\ \bibinfo {pages} {100--112} (\bibinfo {year}
  {2018})}\BibitemShut {NoStop}%
\bibitem [{\citenamefont {Heiss}\ and\ \citenamefont
  {Steeb}(1991)}]{Avoidedlevelcrossings}%
  \BibitemOpen
  \bibfield  {author} {\bibinfo {author} {\bibfnamefont {W.~D.}\ \bibnamefont
  {Heiss}}\ and\ \bibinfo {author} {\bibfnamefont {W.‐H.}\ \bibnamefont
  {Steeb}},\ }\bibfield  {title} {\enquote {\bibinfo {title} {Avoided level
  crossings and riemann sheet structure},}\ }\href {\doibase 10.1063/1.529044}
  {\bibfield  {journal} {\bibinfo  {journal} {Journal of Mathematical Physics}\
  }\textbf {\bibinfo {volume} {32}},\ \bibinfo {pages} {3003--3007} (\bibinfo
  {year} {1991})},\ \Eprint
  {http://arxiv.org/abs/https://doi.org/10.1063/1.529044}
  {https://doi.org/10.1063/1.529044} \BibitemShut {NoStop}%
\bibitem [{\citenamefont {Wiersig}(2019)}]{Nonorthogonalityconstraints}%
  \BibitemOpen
  \bibfield  {author} {\bibinfo {author} {\bibfnamefont {Jan}\ \bibnamefont
  {Wiersig}},\ }\bibfield  {title} {\enquote {\bibinfo {title}
  {Nonorthogonality constraints in open quantum and wave systems},}\ }\href
  {\doibase 10.1103/PhysRevResearch.1.033182} {\bibfield  {journal} {\bibinfo
  {journal} {Phys. Rev. Research}\ }\textbf {\bibinfo {volume} {1}},\ \bibinfo
  {pages} {033182} (\bibinfo {year} {2019})}\BibitemShut {NoStop}%
\bibitem [{\citenamefont {Hodaei}\ \emph {et~al.}(2017)\citenamefont {Hodaei},
  \citenamefont {Hassan}, \citenamefont {Wittek}, \citenamefont
  {Garcia-Gracia}, \citenamefont {El-Ganainy}, \citenamefont
  {Christodoulides},\ and\ \citenamefont {Khajavikhan}}]{hodaei2017enhanced}%
  \BibitemOpen
  \bibfield  {author} {\bibinfo {author} {\bibfnamefont {Hossein}\ \bibnamefont
  {Hodaei}}, \bibinfo {author} {\bibfnamefont {Absar~U}\ \bibnamefont
  {Hassan}}, \bibinfo {author} {\bibfnamefont {Steffen}\ \bibnamefont
  {Wittek}}, \bibinfo {author} {\bibfnamefont {Hipolito}\ \bibnamefont
  {Garcia-Gracia}}, \bibinfo {author} {\bibfnamefont {Ramy}\ \bibnamefont
  {El-Ganainy}}, \bibinfo {author} {\bibfnamefont {Demetrios~N}\ \bibnamefont
  {Christodoulides}}, \ and\ \bibinfo {author} {\bibfnamefont {Mercedeh}\
  \bibnamefont {Khajavikhan}},\ }\bibfield  {title} {\enquote {\bibinfo {title}
  {Enhanced sensitivity at higher-order exceptional points},}\ }\href@noop {}
  {\bibfield  {journal} {\bibinfo  {journal} {Nature}\ }\textbf {\bibinfo
  {volume} {548}},\ \bibinfo {pages} {187--191} (\bibinfo {year}
  {2017})}\BibitemShut {NoStop}%
\bibitem [{\citenamefont {Chen}\ \emph {et~al.}(2017)\citenamefont {Chen},
  \citenamefont {{\"O}zdemir}, \citenamefont {Zhao}, \citenamefont {Wiersig},\
  and\ \citenamefont {Yang}}]{chen2017exceptional}%
  \BibitemOpen
  \bibfield  {author} {\bibinfo {author} {\bibfnamefont {Weijian}\ \bibnamefont
  {Chen}}, \bibinfo {author} {\bibfnamefont {{\c{S}}ahin~Kaya}\ \bibnamefont
  {{\"O}zdemir}}, \bibinfo {author} {\bibfnamefont {Guangming}\ \bibnamefont
  {Zhao}}, \bibinfo {author} {\bibfnamefont {Jan}\ \bibnamefont {Wiersig}}, \
  and\ \bibinfo {author} {\bibfnamefont {Lan}\ \bibnamefont {Yang}},\
  }\bibfield  {title} {\enquote {\bibinfo {title} {Exceptional points enhance
  sensing in an optical microcavity},}\ }\href@noop {} {\bibfield  {journal}
  {\bibinfo  {journal} {Nature}\ }\textbf {\bibinfo {volume} {548}},\ \bibinfo
  {pages} {192--196} (\bibinfo {year} {2017})}\BibitemShut {NoStop}%
\bibitem [{\citenamefont {Frazier}\ and\ \citenamefont
  {Hussein}(2016)}]{frazier2016generalized}%
  \BibitemOpen
  \bibfield  {author} {\bibinfo {author} {\bibfnamefont {Michael~J}\
  \bibnamefont {Frazier}}\ and\ \bibinfo {author} {\bibfnamefont {Mahmoud~I}\
  \bibnamefont {Hussein}},\ }\bibfield  {title} {\enquote {\bibinfo {title}
  {{Generalized Bloch's theorem for viscous metamaterials: Dispersion and
  effective properties based on frequencies and wavenumbers that are
  simultaneously complex}},}\ }\href@noop {} {\bibfield  {journal} {\bibinfo
  {journal} {Comptes Rendus Physique}\ }\textbf {\bibinfo {volume} {17}},\
  \bibinfo {pages} {565--577} (\bibinfo {year} {2016})}\BibitemShut {NoStop}%
\bibitem [{\citenamefont {Zangeneh-Nejad}\ and\ \citenamefont
  {Fleury}(2019)}]{zangeneh2019topological}%
  \BibitemOpen
  \bibfield  {author} {\bibinfo {author} {\bibfnamefont {Farzad}\ \bibnamefont
  {Zangeneh-Nejad}}\ and\ \bibinfo {author} {\bibfnamefont {Romain}\
  \bibnamefont {Fleury}},\ }\bibfield  {title} {\enquote {\bibinfo {title}
  {Topological fano resonances},}\ }\href@noop {} {\bibfield  {journal}
  {\bibinfo  {journal} {Physical review letters}\ }\textbf {\bibinfo {volume}
  {122}},\ \bibinfo {pages} {014301} (\bibinfo {year} {2019})}\BibitemShut
  {NoStop}%
\bibitem [{\citenamefont {Willatzen}\ and\ \citenamefont
  {Christensen}(2014)}]{Willatzen2014}%
  \BibitemOpen
  \bibfield  {author} {\bibinfo {author} {\bibfnamefont {M.}~\bibnamefont
  {Willatzen}}\ and\ \bibinfo {author} {\bibfnamefont {J.}~\bibnamefont
  {Christensen}},\ }\bibfield  {title} {\enquote {\bibinfo {title} {Acoustic
  gain in piezoelectric semiconductors at{$\varepsilon$}-near-zero response},}\
  }\href {\doibase 10.1103/physrevb.89.041201} {\bibfield  {journal} {\bibinfo
  {journal} {Physical Review B}\ }\textbf {\bibinfo {volume} {89}} (\bibinfo
  {year} {2014}),\ 10.1103/physrevb.89.041201}\BibitemShut {NoStop}%
\bibitem [{\citenamefont {Fleury}\ \emph {et~al.}(2015)\citenamefont {Fleury},
  \citenamefont {Sounas},\ and\ \citenamefont {Al{\`{u}}}}]{Fleury2015}%
  \BibitemOpen
  \bibfield  {author} {\bibinfo {author} {\bibfnamefont {Romain}\ \bibnamefont
  {Fleury}}, \bibinfo {author} {\bibfnamefont {Dimitrios}\ \bibnamefont
  {Sounas}}, \ and\ \bibinfo {author} {\bibfnamefont {Andrea}\ \bibnamefont
  {Al{\`{u}}}},\ }\bibfield  {title} {\enquote {\bibinfo {title} {An invisible
  acoustic sensor based on parity-time symmetry},}\ }\href {\doibase
  10.1038/ncomms6905} {\bibfield  {journal} {\bibinfo  {journal} {Nature
  Communications}\ }\textbf {\bibinfo {volume} {6}} (\bibinfo {year} {2015}),\
  10.1038/ncomms6905}\BibitemShut {NoStop}%
\bibitem [{\citenamefont {Aur{\'{e}}gan}\ and\ \citenamefont
  {Pagneux}(2017)}]{Aurgan2017}%
  \BibitemOpen
  \bibfield  {author} {\bibinfo {author} {\bibfnamefont {Yves}\ \bibnamefont
  {Aur{\'{e}}gan}}\ and\ \bibinfo {author} {\bibfnamefont {Vincent}\
  \bibnamefont {Pagneux}},\ }\bibfield  {title} {\enquote {\bibinfo {title}
  {{PT}-symmetric scattering in flow duct acoustics},}\ }\href {\doibase
  10.1103/physrevlett.118.174301} {\bibfield  {journal} {\bibinfo  {journal}
  {Physical Review Letters}\ }\textbf {\bibinfo {volume} {118}} (\bibinfo
  {year} {2017}),\ 10.1103/physrevlett.118.174301}\BibitemShut {NoStop}%
\bibitem [{\citenamefont {Wieczorek}\ \emph {et~al.}(2011)\citenamefont
  {Wieczorek}, \citenamefont {Sensiau}, \citenamefont {Polifke},\ and\
  \citenamefont {Nicoud}}]{thermoacoustic}%
  \BibitemOpen
  \bibfield  {author} {\bibinfo {author} {\bibfnamefont {K.}~\bibnamefont
  {Wieczorek}}, \bibinfo {author} {\bibfnamefont {C.}~\bibnamefont {Sensiau}},
  \bibinfo {author} {\bibfnamefont {W.}~\bibnamefont {Polifke}}, \ and\
  \bibinfo {author} {\bibfnamefont {F.}~\bibnamefont {Nicoud}},\ }\bibfield
  {title} {\enquote {\bibinfo {title} {Assessing non-normal effects in
  thermoacoustic systems with mean flow},}\ }\href {\doibase 10.1063/1.3650418}
  {\bibfield  {journal} {\bibinfo  {journal} {Physics of Fluids}\ }\textbf
  {\bibinfo {volume} {23}},\ \bibinfo {pages} {107103} (\bibinfo {year}
  {2011})},\ \Eprint {http://arxiv.org/abs/https://doi.org/10.1063/1.3650418}
  {https://doi.org/10.1063/1.3650418} \BibitemShut {NoStop}%
\bibitem [{\citenamefont {Achilleos}\ \emph {et~al.}(2017)\citenamefont
  {Achilleos}, \citenamefont {Theocharis}, \citenamefont {Richoux},\ and\
  \citenamefont {Pagneux}}]{Achilleos2017}%
  \BibitemOpen
  \bibfield  {author} {\bibinfo {author} {\bibfnamefont {V.}~\bibnamefont
  {Achilleos}}, \bibinfo {author} {\bibfnamefont {G.}~\bibnamefont
  {Theocharis}}, \bibinfo {author} {\bibfnamefont {O.}~\bibnamefont {Richoux}},
  \ and\ \bibinfo {author} {\bibfnamefont {V.}~\bibnamefont {Pagneux}},\
  }\bibfield  {title} {\enquote {\bibinfo {title} {Non-hermitian acoustic
  metamaterials: Role of exceptional points in sound absorption},}\ }\href
  {\doibase 10.1103/physrevb.95.144303} {\bibfield  {journal} {\bibinfo
  {journal} {Physical Review B}\ }\textbf {\bibinfo {volume} {95}} (\bibinfo
  {year} {2017}),\ 10.1103/physrevb.95.144303}\BibitemShut {NoStop}%
\bibitem [{\citenamefont {Fleury}\ \emph {et~al.}(2014)\citenamefont {Fleury},
  \citenamefont {Sounas},\ and\ \citenamefont {Al{\`{u}}}}]{Fleury2014}%
  \BibitemOpen
  \bibfield  {author} {\bibinfo {author} {\bibfnamefont {Romain}\ \bibnamefont
  {Fleury}}, \bibinfo {author} {\bibfnamefont {Dimitrios~L.}\ \bibnamefont
  {Sounas}}, \ and\ \bibinfo {author} {\bibfnamefont {Andrea}\ \bibnamefont
  {Al{\`{u}}}},\ }\bibfield  {title} {\enquote {\bibinfo {title} {Negative
  refraction and planar focusing based on parity-time symmetric
  metasurfaces},}\ }\href {\doibase 10.1103/physrevlett.113.023903} {\bibfield
  {journal} {\bibinfo  {journal} {Physical Review Letters}\ }\textbf {\bibinfo
  {volume} {113}} (\bibinfo {year} {2014}),\
  10.1103/physrevlett.113.023903}\BibitemShut {NoStop}%
\bibitem [{\citenamefont {El-Ganainy}\ \emph {et~al.}(2018)\citenamefont
  {El-Ganainy}, \citenamefont {Makris}, \citenamefont {Khajavikhan},
  \citenamefont {Musslimani}, \citenamefont {Rotter},\ and\ \citenamefont
  {Christodoulides}}]{ElGanainy2018}%
  \BibitemOpen
  \bibfield  {author} {\bibinfo {author} {\bibfnamefont {Ramy}\ \bibnamefont
  {El-Ganainy}}, \bibinfo {author} {\bibfnamefont {Konstantinos~G.}\
  \bibnamefont {Makris}}, \bibinfo {author} {\bibfnamefont {Mercedeh}\
  \bibnamefont {Khajavikhan}}, \bibinfo {author} {\bibfnamefont {Ziad~H.}\
  \bibnamefont {Musslimani}}, \bibinfo {author} {\bibfnamefont {Stefan}\
  \bibnamefont {Rotter}}, \ and\ \bibinfo {author} {\bibfnamefont
  {Demetrios~N.}\ \bibnamefont {Christodoulides}},\ }\bibfield  {title}
  {\enquote {\bibinfo {title} {Non-hermitian physics and {PT} symmetry},}\
  }\href {\doibase 10.1038/nphys4323} {\bibfield  {journal} {\bibinfo
  {journal} {Nature Physics}\ }\textbf {\bibinfo {volume} {14}},\ \bibinfo
  {pages} {11--19} (\bibinfo {year} {2018})}\BibitemShut {NoStop}%
\bibitem [{\citenamefont {Monticone}\ \emph {et~al.}(2016)\citenamefont
  {Monticone}, \citenamefont {Valagiannopoulos},\ and\ \citenamefont
  {Al{\`{u}}}}]{Monticone2016}%
  \BibitemOpen
  \bibfield  {author} {\bibinfo {author} {\bibfnamefont {Francesco}\
  \bibnamefont {Monticone}}, \bibinfo {author} {\bibfnamefont
  {Constantinos~A.}\ \bibnamefont {Valagiannopoulos}}, \ and\ \bibinfo {author}
  {\bibfnamefont {Andrea}\ \bibnamefont {Al{\`{u}}}},\ }\bibfield  {title}
  {\enquote {\bibinfo {title} {Parity-time symmetric nonlocal metasurfaces:
  All-angle negative refraction and volumetric imaging},}\ }\href {\doibase
  10.1103/physrevx.6.041018} {\bibfield  {journal} {\bibinfo  {journal}
  {Physical Review X}\ }\textbf {\bibinfo {volume} {6}} (\bibinfo {year}
  {2016}),\ 10.1103/physrevx.6.041018}\BibitemShut {NoStop}%
\bibitem [{\citenamefont {Konotop}\ \emph {et~al.}(2016)\citenamefont
  {Konotop}, \citenamefont {Yang},\ and\ \citenamefont
  {Zezyulin}}]{Konotop2016}%
  \BibitemOpen
  \bibfield  {author} {\bibinfo {author} {\bibfnamefont {Vladimir~V.}\
  \bibnamefont {Konotop}}, \bibinfo {author} {\bibfnamefont {Jianke}\
  \bibnamefont {Yang}}, \ and\ \bibinfo {author} {\bibfnamefont {Dmitry~A.}\
  \bibnamefont {Zezyulin}},\ }\bibfield  {title} {\enquote {\bibinfo {title}
  {Nonlinear waves {inPT}-symmetric systems},}\ }\href {\doibase
  10.1103/revmodphys.88.035002} {\bibfield  {journal} {\bibinfo  {journal}
  {Reviews of Modern Physics}\ }\textbf {\bibinfo {volume} {88}} (\bibinfo
  {year} {2016}),\ 10.1103/revmodphys.88.035002}\BibitemShut {NoStop}%
\bibitem [{\citenamefont {Song}\ \emph {et~al.}(2014)\citenamefont {Song},
  \citenamefont {Bai}, \citenamefont {Hang},\ and\ \citenamefont
  {Lai}}]{Song2014}%
  \BibitemOpen
  \bibfield  {author} {\bibinfo {author} {\bibfnamefont {J~Z}\ \bibnamefont
  {Song}}, \bibinfo {author} {\bibfnamefont {P}~\bibnamefont {Bai}}, \bibinfo
  {author} {\bibfnamefont {Z~H}\ \bibnamefont {Hang}}, \ and\ \bibinfo {author}
  {\bibfnamefont {Yun}\ \bibnamefont {Lai}},\ }\bibfield  {title} {\enquote
  {\bibinfo {title} {Acoustic coherent perfect absorbers},}\ }\href {\doibase
  10.1088/1367-2630/16/3/033026} {\bibfield  {journal} {\bibinfo  {journal}
  {New Journal of Physics}\ }\textbf {\bibinfo {volume} {16}},\ \bibinfo
  {pages} {033026} (\bibinfo {year} {2014})}\BibitemShut {NoStop}%
\bibitem [{\citenamefont {Wei}\ \emph {et~al.}(2014)\citenamefont {Wei},
  \citenamefont {Croënne}, \citenamefont {Chu},\ and\ \citenamefont
  {Li}}]{Wei2014}%
  \BibitemOpen
  \bibfield  {author} {\bibinfo {author} {\bibfnamefont {Pengjiang}\
  \bibnamefont {Wei}}, \bibinfo {author} {\bibfnamefont {Charles}\ \bibnamefont
  {Croënne}}, \bibinfo {author} {\bibfnamefont {Sai~Tak}\ \bibnamefont {Chu}},
  \ and\ \bibinfo {author} {\bibfnamefont {Jensen}\ \bibnamefont {Li}},\
  }\bibfield  {title} {\enquote {\bibinfo {title} {Symmetrical and
  anti-symmetrical coherent perfect absorption for acoustic waves},}\ }\href
  {\doibase 10.1063/1.4869462} {\bibfield  {journal} {\bibinfo  {journal}
  {Applied Physics Letters}\ }\textbf {\bibinfo {volume} {104}},\ \bibinfo
  {pages} {121902} (\bibinfo {year} {2014})}\BibitemShut {NoStop}%
\bibitem [{\citenamefont {Feshbach}(1958)}]{FESHBACH1958357}%
  \BibitemOpen
  \bibfield  {author} {\bibinfo {author} {\bibfnamefont {Herman}\ \bibnamefont
  {Feshbach}},\ }\bibfield  {title} {\enquote {\bibinfo {title} {Unified theory
  of nuclear reactions},}\ }\href {\doibase
  https://doi.org/10.1016/0003-4916(58)90007-1} {\bibfield  {journal} {\bibinfo
   {journal} {Annals of Physics}\ }\textbf {\bibinfo {volume} {5}},\ \bibinfo
  {pages} {357--390} (\bibinfo {year} {1958})}\BibitemShut {NoStop}%
\bibitem [{\citenamefont {Feshbach}(1962)}]{Feshbach1962}%
  \BibitemOpen
  \bibfield  {author} {\bibinfo {author} {\bibfnamefont {Herman}\ \bibnamefont
  {Feshbach}},\ }\bibfield  {title} {\enquote {\bibinfo {title} {A unified
  theory of nuclear reactions. {II}},}\ }\href {\doibase
  10.1016/0003-4916(62)90221-x} {\bibfield  {journal} {\bibinfo  {journal}
  {Annals of Physics}\ }\textbf {\bibinfo {volume} {19}},\ \bibinfo {pages}
  {287--313} (\bibinfo {year} {1962})}\BibitemShut {NoStop}%
\bibitem [{\citenamefont {Deymier}\ and\ \citenamefont
  {Runge}(2017)}]{Deymier2017}%
  \BibitemOpen
  \bibfield  {author} {\bibinfo {author} {\bibfnamefont {Pierre}\ \bibnamefont
  {Deymier}}\ and\ \bibinfo {author} {\bibfnamefont {Keith}\ \bibnamefont
  {Runge}},\ }\href {\doibase 10.1007/978-3-319-62380-1} {\emph {\bibinfo
  {title} {Sound Topology, Duality, Coherence and Wave-Mixing}}}\ (\bibinfo
  {publisher} {Springer International Publishing},\ \bibinfo {year}
  {2017})\BibitemShut {NoStop}%
\bibitem [{\citenamefont {Sadasivam}\ \emph {et~al.}(2014)\citenamefont
  {Sadasivam}, \citenamefont {Che}, \citenamefont {Huang}, \citenamefont
  {Chen}, \citenamefont {Kumar},\ and\ \citenamefont
  {Fisher}}]{sadasivam2014atomistic}%
  \BibitemOpen
  \bibfield  {author} {\bibinfo {author} {\bibfnamefont {Sridhar}\ \bibnamefont
  {Sadasivam}}, \bibinfo {author} {\bibfnamefont {Yuhang}\ \bibnamefont {Che}},
  \bibinfo {author} {\bibfnamefont {Zhen}\ \bibnamefont {Huang}}, \bibinfo
  {author} {\bibfnamefont {Liang}\ \bibnamefont {Chen}}, \bibinfo {author}
  {\bibfnamefont {Satish}\ \bibnamefont {Kumar}}, \ and\ \bibinfo {author}
  {\bibfnamefont {Timothy~S}\ \bibnamefont {Fisher}},\ }\bibfield  {title}
  {\enquote {\bibinfo {title} {The atomistic green's function method for
  interfacial phonon transport},}\ }\href@noop {} {\bibfield  {journal}
  {\bibinfo  {journal} {Annual Review of Heat Transfer}\ }\textbf {\bibinfo
  {volume} {17}} (\bibinfo {year} {2014})}\BibitemShut {NoStop}%
\bibitem [{\citenamefont {Mingo}\ and\ \citenamefont
  {Yang}(2003)}]{mingo2003phonon}%
  \BibitemOpen
  \bibfield  {author} {\bibinfo {author} {\bibfnamefont {N}~\bibnamefont
  {Mingo}}\ and\ \bibinfo {author} {\bibfnamefont {Liu}\ \bibnamefont {Yang}},\
  }\bibfield  {title} {\enquote {\bibinfo {title} {Phonon transport in
  nanowires coated with an amorphous material: An atomistic green’s function
  approach},}\ }\href@noop {} {\bibfield  {journal} {\bibinfo  {journal}
  {Physical Review B}\ }\textbf {\bibinfo {volume} {68}},\ \bibinfo {pages}
  {245406} (\bibinfo {year} {2003})}\BibitemShut {NoStop}%
\bibitem [{\citenamefont {Khalatnikov}(2018)}]{khalatnikov2018heat}%
  \BibitemOpen
  \bibfield  {author} {\bibinfo {author} {\bibfnamefont {IM}~\bibnamefont
  {Khalatnikov}},\ }\bibfield  {title} {\enquote {\bibinfo {title} {Heat
  exchange between a solid and helium ii},}\ }in\ \href@noop {} {\emph
  {\bibinfo {booktitle} {An introduction to the theory of superfluidity}}}\
  (\bibinfo  {publisher} {CRC Press},\ \bibinfo {year} {2018})\ pp.\ \bibinfo
  {pages} {138--146}\BibitemShut {NoStop}%
\bibitem [{\citenamefont {Little}(1959)}]{little1959transport}%
  \BibitemOpen
  \bibfield  {author} {\bibinfo {author} {\bibfnamefont {WA}~\bibnamefont
  {Little}},\ }\bibfield  {title} {\enquote {\bibinfo {title} {The transport of
  heat between dissimilar solids at low temperatures},}\ }\href@noop {}
  {\bibfield  {journal} {\bibinfo  {journal} {Canadian Journal of Physics}\
  }\textbf {\bibinfo {volume} {37}},\ \bibinfo {pages} {334--349} (\bibinfo
  {year} {1959})}\BibitemShut {NoStop}%
\bibitem [{\citenamefont {Swartz}\ and\ \citenamefont
  {Pohl}(1989)}]{swartz1989thermal}%
  \BibitemOpen
  \bibfield  {author} {\bibinfo {author} {\bibfnamefont {Eric~T}\ \bibnamefont
  {Swartz}}\ and\ \bibinfo {author} {\bibfnamefont {Robert~O}\ \bibnamefont
  {Pohl}},\ }\bibfield  {title} {\enquote {\bibinfo {title} {Thermal boundary
  resistance},}\ }\href@noop {} {\bibfield  {journal} {\bibinfo  {journal}
  {Reviews of modern physics}\ }\textbf {\bibinfo {volume} {61}},\ \bibinfo
  {pages} {605} (\bibinfo {year} {1989})}\BibitemShut {NoStop}%
\bibitem [{\citenamefont {Datta}(2005)}]{datta2005quantum}%
  \BibitemOpen
  \bibfield  {author} {\bibinfo {author} {\bibfnamefont {Supriyo}\ \bibnamefont
  {Datta}},\ }\href@noop {} {\emph {\bibinfo {title} {Quantum transport: atom
  to transistor}}}\ (\bibinfo  {publisher} {Cambridge university press},\
  \bibinfo {year} {2005})\BibitemShut {NoStop}%
\bibitem [{\citenamefont {Hopkins}\ \emph {et~al.}(2009)\citenamefont
  {Hopkins}, \citenamefont {Norris}, \citenamefont {Tsegaye},\ and\
  \citenamefont {Ghosh}}]{nano}%
  \BibitemOpen
  \bibfield  {author} {\bibinfo {author} {\bibfnamefont {Patrick~E.}\
  \bibnamefont {Hopkins}}, \bibinfo {author} {\bibfnamefont {Pamela~M.}\
  \bibnamefont {Norris}}, \bibinfo {author} {\bibfnamefont {Mikiyas~S.}\
  \bibnamefont {Tsegaye}}, \ and\ \bibinfo {author} {\bibfnamefont {Avik~W.}\
  \bibnamefont {Ghosh}},\ }\bibfield  {title} {\enquote {\bibinfo {title}
  {Extracting phonon thermal conductance across atomic junctions:
  Nonequilibrium green’s function approach compared to semiclassical
  methods},}\ }\href {\doibase 10.1063/1.3212974} {\bibfield  {journal}
  {\bibinfo  {journal} {Journal of Applied Physics}\ }\textbf {\bibinfo
  {volume} {106}},\ \bibinfo {pages} {063503} (\bibinfo {year} {2009})},\
  \Eprint {http://arxiv.org/abs/https://doi.org/10.1063/1.3212974}
  {https://doi.org/10.1063/1.3212974} \BibitemShut {NoStop}%
\bibitem [{\citenamefont {Khomyakov}\ \emph {et~al.}(2005)\citenamefont
  {Khomyakov}, \citenamefont {Brocks}, \citenamefont {Karpan}, \citenamefont
  {Zwierzycki},\ and\ \citenamefont {Kelly}}]{PhysRevB.72.035450}%
  \BibitemOpen
  \bibfield  {author} {\bibinfo {author} {\bibfnamefont {P.~A.}\ \bibnamefont
  {Khomyakov}}, \bibinfo {author} {\bibfnamefont {G.}~\bibnamefont {Brocks}},
  \bibinfo {author} {\bibfnamefont {V.}~\bibnamefont {Karpan}}, \bibinfo
  {author} {\bibfnamefont {M.}~\bibnamefont {Zwierzycki}}, \ and\ \bibinfo
  {author} {\bibfnamefont {P.~J.}\ \bibnamefont {Kelly}},\ }\bibfield  {title}
  {\enquote {\bibinfo {title} {Conductance calculations for quantum wires and
  interfaces: Mode matching and green's functions},}\ }\href {\doibase
  10.1103/PhysRevB.72.035450} {\bibfield  {journal} {\bibinfo  {journal} {Phys.
  Rev. B}\ }\textbf {\bibinfo {volume} {72}},\ \bibinfo {pages} {035450}
  (\bibinfo {year} {2005})}\BibitemShut {NoStop}%
\bibitem [{\citenamefont {Ando}(1991)}]{PhysRevB.44.8017}%
  \BibitemOpen
  \bibfield  {author} {\bibinfo {author} {\bibfnamefont {T.}~\bibnamefont
  {Ando}},\ }\bibfield  {title} {\enquote {\bibinfo {title} {Quantum point
  contacts in magnetic fields},}\ }\href {\doibase 10.1103/PhysRevB.44.8017}
  {\bibfield  {journal} {\bibinfo  {journal} {Phys. Rev. B}\ }\textbf {\bibinfo
  {volume} {44}},\ \bibinfo {pages} {8017--8027} (\bibinfo {year}
  {1991})}\BibitemShut {NoStop}%
\bibitem [{\citenamefont {Flax}\ \emph {et~al.}(1978)\citenamefont {Flax},
  \citenamefont {Dragonette},\ and\ \citenamefont
  {{\"U}berall}}]{flax1978theory}%
  \BibitemOpen
  \bibfield  {author} {\bibinfo {author} {\bibfnamefont {L}~\bibnamefont
  {Flax}}, \bibinfo {author} {\bibfnamefont {LR}~\bibnamefont {Dragonette}}, \
  and\ \bibinfo {author} {\bibfnamefont {H}~\bibnamefont {{\"U}berall}},\
  }\bibfield  {title} {\enquote {\bibinfo {title} {Theory of elastic resonance
  excitation by sound scattering},}\ }\href@noop {} {\bibfield  {journal}
  {\bibinfo  {journal} {The Journal of the Acoustical Society of America}\
  }\textbf {\bibinfo {volume} {63}},\ \bibinfo {pages} {723--731} (\bibinfo
  {year} {1978})}\BibitemShut {NoStop}%
\bibitem [{\citenamefont {Neubauer}\ \emph {et~al.}(1969)\citenamefont
  {Neubauer}, \citenamefont {Ugin{\v{c}}ius},\ and\ \citenamefont
  {{\"U}berall}}]{neubauer1969theory}%
  \BibitemOpen
  \bibfield  {author} {\bibinfo {author} {\bibfnamefont {WG}~\bibnamefont
  {Neubauer}}, \bibinfo {author} {\bibfnamefont {P}~\bibnamefont
  {Ugin{\v{c}}ius}}, \ and\ \bibinfo {author} {\bibfnamefont {H}~\bibnamefont
  {{\"U}berall}},\ }\bibfield  {title} {\enquote {\bibinfo {title} {Theory of
  creeping waves in acoustics and their experimental demonstration},}\
  }\href@noop {} {\bibfield  {journal} {\bibinfo  {journal} {Zeitschrift
  f{\"u}r Naturforschung A}\ }\textbf {\bibinfo {volume} {24}},\ \bibinfo
  {pages} {691--700} (\bibinfo {year} {1969})}\BibitemShut {NoStop}%
\bibitem [{\citenamefont {{\"U}berall}\ \emph {et~al.}(1977)\citenamefont
  {{\"U}berall}, \citenamefont {Dragonette},\ and\ \citenamefont
  {Flax}}]{uberall1977relation}%
  \BibitemOpen
  \bibfield  {author} {\bibinfo {author} {\bibfnamefont {H}~\bibnamefont
  {{\"U}berall}}, \bibinfo {author} {\bibfnamefont {LR}~\bibnamefont
  {Dragonette}}, \ and\ \bibinfo {author} {\bibfnamefont {L}~\bibnamefont
  {Flax}},\ }\bibfield  {title} {\enquote {\bibinfo {title} {Relation between
  creeping waves and normal modes of vibration of a curved body},}\ }\href@noop
  {} {\bibfield  {journal} {\bibinfo  {journal} {The Journal of the Acoustical
  Society of America}\ }\textbf {\bibinfo {volume} {61}},\ \bibinfo {pages}
  {711--715} (\bibinfo {year} {1977})}\BibitemShut {NoStop}%
\bibitem [{\citenamefont {Hackman}\ and\ \citenamefont
  {Sammelmann}(1989)}]{hackman1989existence}%
  \BibitemOpen
  \bibfield  {author} {\bibinfo {author} {\bibfnamefont {Roger~H}\ \bibnamefont
  {Hackman}}\ and\ \bibinfo {author} {\bibfnamefont {Gary~S}\ \bibnamefont
  {Sammelmann}},\ }\bibfield  {title} {\enquote {\bibinfo {title} {On the
  existence of the rayleigh wave dipole resonance},}\ }\href@noop {} {\bibfield
   {journal} {\bibinfo  {journal} {The Journal of the Acoustical Society of
  America}\ }\textbf {\bibinfo {volume} {85}},\ \bibinfo {pages} {2284--2289}
  (\bibinfo {year} {1989})}\BibitemShut {NoStop}%
\bibitem [{\citenamefont {Hackman}(1993)}]{hackman1993acoustic}%
  \BibitemOpen
  \bibfield  {author} {\bibinfo {author} {\bibfnamefont {Roger~H}\ \bibnamefont
  {Hackman}},\ }\bibfield  {title} {\enquote {\bibinfo {title} {Acoustic
  scattering from elastic solids},}\ }in\ \href@noop {} {\emph {\bibinfo
  {booktitle} {Physical acoustics}}},\ Vol.~\bibinfo {volume} {22}\ (\bibinfo
  {publisher} {Elsevier},\ \bibinfo {year} {1993})\ pp.\ \bibinfo {pages}
  {1--194}\BibitemShut {NoStop}%
\bibitem [{\citenamefont {Klaiman}\ and\ \citenamefont
  {Moiseyev}(2010)}]{klaiman2010absolute}%
  \BibitemOpen
  \bibfield  {author} {\bibinfo {author} {\bibfnamefont {Shachar}\ \bibnamefont
  {Klaiman}}\ and\ \bibinfo {author} {\bibfnamefont {Nimrod}\ \bibnamefont
  {Moiseyev}},\ }\bibfield  {title} {\enquote {\bibinfo {title} {The absolute
  position of a resonance peak},}\ }\href@noop {} {\bibfield  {journal}
  {\bibinfo  {journal} {Journal of Physics B: Atomic, Molecular and Optical
  Physics}\ }\textbf {\bibinfo {volume} {43}},\ \bibinfo {pages} {185205}
  (\bibinfo {year} {2010})}\BibitemShut {NoStop}%
\bibitem [{\citenamefont {Kittel}\ and\ \citenamefont
  {McEuen}(1976)}]{kittel1976introduction}%
  \BibitemOpen
  \bibfield  {author} {\bibinfo {author} {\bibfnamefont {Charles}\ \bibnamefont
  {Kittel}}\ and\ \bibinfo {author} {\bibfnamefont {Paul}\ \bibnamefont
  {McEuen}},\ }\href@noop {} {\emph {\bibinfo {title} {{Introduction to solid
  state physics}}}},\ Vol.~\bibinfo {volume} {8}\ (\bibinfo  {publisher} {Wiley
  New York},\ \bibinfo {year} {1976})\BibitemShut {NoStop}%
\bibitem [{\citenamefont {Garmon}\ \emph
  {et~al.}(2015{\natexlab{b}})\citenamefont {Garmon}, \citenamefont
  {Gianfreda},\ and\ \citenamefont {Hatano}}]{garmon2015bound}%
  \BibitemOpen
  \bibfield  {author} {\bibinfo {author} {\bibfnamefont {Savannah}\
  \bibnamefont {Garmon}}, \bibinfo {author} {\bibfnamefont {Mariagiovanna}\
  \bibnamefont {Gianfreda}}, \ and\ \bibinfo {author} {\bibfnamefont
  {Naomichi}\ \bibnamefont {Hatano}},\ }\bibfield  {title} {\enquote {\bibinfo
  {title} {Bound states, scattering states, and resonant states in pt-symmetric
  open quantum systems},}\ }\href@noop {} {\bibfield  {journal} {\bibinfo
  {journal} {Physical Review A}\ }\textbf {\bibinfo {volume} {92}},\ \bibinfo
  {pages} {022125} (\bibinfo {year} {2015}{\natexlab{b}})}\BibitemShut
  {NoStop}%
\bibitem [{\citenamefont {Sasada}\ and\ \citenamefont
  {Hatano}(2008)}]{sasada2008calculation}%
  \BibitemOpen
  \bibfield  {author} {\bibinfo {author} {\bibfnamefont {Keita}\ \bibnamefont
  {Sasada}}\ and\ \bibinfo {author} {\bibfnamefont {Naomichi}\ \bibnamefont
  {Hatano}},\ }\bibfield  {title} {\enquote {\bibinfo {title} {Calculation of
  the self-energy of open quantum systems},}\ }\href@noop {} {\bibfield
  {journal} {\bibinfo  {journal} {Journal of the Physical Society of Japan}\
  }\textbf {\bibinfo {volume} {77}},\ \bibinfo {pages} {025003--025003}
  (\bibinfo {year} {2008})}\BibitemShut {NoStop}%
\bibitem [{\citenamefont {Siegert}(1939)}]{siegert1939derivation}%
  \BibitemOpen
  \bibfield  {author} {\bibinfo {author} {\bibfnamefont {Arnold~JF}\
  \bibnamefont {Siegert}},\ }\bibfield  {title} {\enquote {\bibinfo {title} {On
  the derivation of the dispersion formula for nuclear reactions},}\
  }\href@noop {} {\bibfield  {journal} {\bibinfo  {journal} {Physical Review}\
  }\textbf {\bibinfo {volume} {56}},\ \bibinfo {pages} {750} (\bibinfo {year}
  {1939})}\BibitemShut {NoStop}%
\bibitem [{\citenamefont {Hatano}(2013)}]{hatano2013equivalence}%
  \BibitemOpen
  \bibfield  {author} {\bibinfo {author} {\bibfnamefont {Naomichi}\
  \bibnamefont {Hatano}},\ }\bibfield  {title} {\enquote {\bibinfo {title}
  {Equivalence of the effective hamiltonian approach and the siegert boundary
  condition for resonant states},}\ }\href@noop {} {\bibfield  {journal}
  {\bibinfo  {journal} {Fortschritte der Physik}\ }\textbf {\bibinfo {volume}
  {61}},\ \bibinfo {pages} {238--249} (\bibinfo {year} {2013})}\BibitemShut
  {NoStop}%
\bibitem [{\citenamefont {Sasada}\ \emph {et~al.}(2011)\citenamefont {Sasada},
  \citenamefont {Hatano},\ and\ \citenamefont {Ordonez}}]{sasada2011resonant}%
  \BibitemOpen
  \bibfield  {author} {\bibinfo {author} {\bibfnamefont {Keita}\ \bibnamefont
  {Sasada}}, \bibinfo {author} {\bibfnamefont {Naomichi}\ \bibnamefont
  {Hatano}}, \ and\ \bibinfo {author} {\bibfnamefont {Gonzalo}\ \bibnamefont
  {Ordonez}},\ }\bibfield  {title} {\enquote {\bibinfo {title} {Resonant
  spectrum analysis of the conductance of an open quantum system and three
  types of fano parameter},}\ }\href@noop {} {\bibfield  {journal} {\bibinfo
  {journal} {Journal of the Physical Society of Japan}\ }\textbf {\bibinfo
  {volume} {80}},\ \bibinfo {pages} {104707} (\bibinfo {year}
  {2011})}\BibitemShut {NoStop}%
\bibitem [{\citenamefont {Tisseur}\ and\ \citenamefont
  {Meerbergen}(2001)}]{tisseur2001quadratic}%
  \BibitemOpen
  \bibfield  {author} {\bibinfo {author} {\bibfnamefont {Fran{\c{c}}oise}\
  \bibnamefont {Tisseur}}\ and\ \bibinfo {author} {\bibfnamefont {Karl}\
  \bibnamefont {Meerbergen}},\ }\bibfield  {title} {\enquote {\bibinfo {title}
  {The quadratic eigenvalue problem},}\ }\href@noop {} {\bibfield  {journal}
  {\bibinfo  {journal} {SIAM review}\ }\textbf {\bibinfo {volume} {43}},\
  \bibinfo {pages} {235--286} (\bibinfo {year} {2001})}\BibitemShut {NoStop}%
\bibitem [{\citenamefont {Fisher}\ and\ \citenamefont
  {Lee}(1981)}]{fisher1981relation}%
  \BibitemOpen
  \bibfield  {author} {\bibinfo {author} {\bibfnamefont {Daniel~S}\
  \bibnamefont {Fisher}}\ and\ \bibinfo {author} {\bibfnamefont {Patrick~A}\
  \bibnamefont {Lee}},\ }\bibfield  {title} {\enquote {\bibinfo {title}
  {Relation between conductivity and transmission matrix},}\ }\href@noop {}
  {\bibfield  {journal} {\bibinfo  {journal} {Physical Review B}\ }\textbf
  {\bibinfo {volume} {23}},\ \bibinfo {pages} {6851} (\bibinfo {year}
  {1981})}\BibitemShut {NoStop}%
\bibitem [{\citenamefont {Hatano}\ \emph {et~al.}(2008)\citenamefont {Hatano},
  \citenamefont {Sasada}, \citenamefont {Nakamura},\ and\ \citenamefont
  {Petrosky}}]{hatano2008some}%
  \BibitemOpen
  \bibfield  {author} {\bibinfo {author} {\bibfnamefont {Naomichi}\
  \bibnamefont {Hatano}}, \bibinfo {author} {\bibfnamefont {Keita}\
  \bibnamefont {Sasada}}, \bibinfo {author} {\bibfnamefont {Hiroaki}\
  \bibnamefont {Nakamura}}, \ and\ \bibinfo {author} {\bibfnamefont {Tomio}\
  \bibnamefont {Petrosky}},\ }\bibfield  {title} {\enquote {\bibinfo {title}
  {Some properties of the resonant state in quantum mechanics and its
  computation},}\ }\href@noop {} {\bibfield  {journal} {\bibinfo  {journal}
  {Progress of theoretical physics}\ }\textbf {\bibinfo {volume} {119}},\
  \bibinfo {pages} {187--222} (\bibinfo {year} {2008})}\BibitemShut {NoStop}%
\bibitem [{\citenamefont {Rotter}(2009)}]{Rotter_2009}%
  \BibitemOpen
  \bibfield  {author} {\bibinfo {author} {\bibfnamefont {Ingrid}\ \bibnamefont
  {Rotter}},\ }\bibfield  {title} {\enquote {\bibinfo {title} {A non-hermitian
  hamilton operator and the physics of open quantum systems},}\ }\href
  {\doibase 10.1088/1751-8113/42/15/153001} {\bibfield  {journal} {\bibinfo
  {journal} {Journal of Physics A: Mathematical and Theoretical}\ }\textbf
  {\bibinfo {volume} {42}},\ \bibinfo {pages} {153001} (\bibinfo {year}
  {2009})}\BibitemShut {NoStop}%
\bibitem [{\citenamefont {Wang}\ \emph {et~al.}(2021)\citenamefont {Wang},
  \citenamefont {Mokhtari}, \citenamefont {Srivastava},\ and\ \citenamefont
  {Amirkhizi}}]{wang2021angledependent}%
  \BibitemOpen
  \bibfield  {author} {\bibinfo {author} {\bibfnamefont {Weidi}\ \bibnamefont
  {Wang}}, \bibinfo {author} {\bibfnamefont {Amir~Ashkan}\ \bibnamefont
  {Mokhtari}}, \bibinfo {author} {\bibfnamefont {Ankit}\ \bibnamefont
  {Srivastava}}, \ and\ \bibinfo {author} {\bibfnamefont {Alireza~V.}\
  \bibnamefont {Amirkhizi}},\ }\href@noop {} {\enquote {\bibinfo {title}
  {Angle-dependent phononic dynamics for deep learning and source
  localization},}\ } (\bibinfo {year} {2021}),\ \Eprint
  {http://arxiv.org/abs/2108.12080} {arXiv:2108.12080 [physics.app-ph]}
  \BibitemShut {NoStop}%
\bibitem [{\citenamefont {Scandrett}(2002)}]{scandrett2002scattering}%
  \BibitemOpen
  \bibfield  {author} {\bibinfo {author} {\bibfnamefont {Clyde}\ \bibnamefont
  {Scandrett}},\ }\bibfield  {title} {\enquote {\bibinfo {title} {Scattering
  and active acoustic control from a submerged spherical shell},}\ }\href@noop
  {} {\bibfield  {journal} {\bibinfo  {journal} {The Journal of the Acoustical
  Society of America}\ }\textbf {\bibinfo {volume} {111}},\ \bibinfo {pages}
  {893--907} (\bibinfo {year} {2002})}\BibitemShut {NoStop}%
\bibitem [{\citenamefont {Hu}\ \emph {et~al.}(1954)\citenamefont {Hu} \emph
  {et~al.}}]{hu1954general}%
  \BibitemOpen
  \bibfield  {author} {\bibinfo {author} {\bibfnamefont {Hai-Chang}\
  \bibnamefont {Hu}} \emph {et~al.},\ }\bibfield  {title} {\enquote {\bibinfo
  {title} {On the general theory of elasticity for a spherically isotropic
  medium},}\ }\href@noop {} {\bibfield  {journal} {\bibinfo  {journal}
  {Scientia Sinica}\ }\textbf {\bibinfo {volume} {3}},\ \bibinfo {pages}
  {247--260} (\bibinfo {year} {1954})}\BibitemShut {NoStop}%
\bibitem [{\citenamefont {Ding}\ and\ \citenamefont
  {Chen}(1996)}]{ding1996natural}%
  \BibitemOpen
  \bibfield  {author} {\bibinfo {author} {\bibfnamefont {HJ}~\bibnamefont
  {Ding}}\ and\ \bibinfo {author} {\bibfnamefont {WQ}~\bibnamefont {Chen}},\
  }\bibfield  {title} {\enquote {\bibinfo {title} {Natural frequencies of an
  elastic spherically isotropic hollow sphere submerged in a compressible fluid
  medium},}\ }\href@noop {} {\bibfield  {journal} {\bibinfo  {journal} {Journal
  of sound and vibration}\ }\textbf {\bibinfo {volume} {192}},\ \bibinfo
  {pages} {173--198} (\bibinfo {year} {1996})}\BibitemShut {NoStop}%
\bibitem [{\citenamefont {Erdelyi}\ \emph {et~al.}(1972)\citenamefont
  {Erdelyi}, \citenamefont {Magnus},\ and\ \citenamefont
  {Oberhettinger}}]{mathhandbook}%
  \BibitemOpen
  \bibfield  {author} {\bibinfo {author} {\bibfnamefont {A}~\bibnamefont
  {Erdelyi}}, \bibinfo {author} {\bibfnamefont {W}~\bibnamefont {Magnus}}, \
  and\ \bibinfo {author} {\bibfnamefont {F}~\bibnamefont {Oberhettinger}},\
  }\href@noop {} {\enquote {\bibinfo {title} {M. abramowitz and ia stegun,
  handbook of mathematical functions},}\ } (\bibinfo {year} {1972})\BibitemShut
  {NoStop}%
\end{thebibliography}
%
\end{document}